\documentclass[11pt, a4paper]{article}

\usepackage{jheparxiv}
\usepackage[utf8]{inputenc}

\usepackage{amsmath,amssymb,graphicx}
\usepackage{mathrsfs}
\usepackage{color}
\usepackage{hyperref}
\usepackage{slashed}
\usepackage[normalem]{ulem} %% for strikethrough text
\usepackage{tikz}
\usepackage{verbatim}

\newcommand{\be}{\begin{equation}}
\newcommand{\ee}{\end{equation}}
\newcommand{\bi}{\begin{itemize}}
\newcommand{\ei}{\end{itemize}}
\newcommand{\bea}{\begin{eqnarray}}
\newcommand{\eea}{\end{eqnarray}}
\newcommand{\tr}{\text{tr}\,}

\newcommand{\D}{\mathrm{D}}
\newcommand{\cE}{\mathcal{E}}
\newcommand{\im}{\mathrm{i}}
\renewcommand{\d}{\mathrm{d}}
\newcommand{\la}{\langle}
\newcommand{\ra}{\rangle}

\newcommand{\ud}{\mathrm{d}}
\newcommand{\LCm}{{\scriptscriptstyle -}} %LC supersripts
\newcommand{\LCp}{{\scriptscriptstyle +}}

\newcommand{\LCperp}{{\scriptscriptstyle \perp}}

\newcommand{\sa}{\mathsf{a}}
\newcommand{\e}{\mathrm{e}}
\newcommand{\cA}{\mathcal{A}}

\newcommand{\scG}{\mathscr{G}}
\newcommand{\C}{\mathbb{C}}

\begin{document}

\subheader{\hfill \texttt{IMPERIAL-TP-TA-2019-01}}

\title{Gluon helicity flip in a plane wave background}

\author[1]{Tim Adamo}
\author[2]{\& Anton Ilderton}
\emailAdd{t.adamo@imperial.ac.uk}
\emailAdd{anton.ilderton@plymouth.ac.uk}

\affiliation[1]{Theoretical Physics Group, Blackett Laboratory \\
        Imperial College London, SW7 2AZ, United Kingdom}
\affiliation[2]{Centre for Mathematical Sciences \\
        University of Plymouth, PL4 8AA, United Kingdom}

\abstract{We compute the leading probability for a gluon to flip helicity state
upon traversing a background plane wave gauge field in pure Yang-Mills
theory and QCD, with an arbitrary number of colours and flavours. This
is a one-loop calculation in perturbative gauge theory around the
gluonic plane wave background, which is treated without approximation
(i.e., to all orders in the coupling). We introduce a background--dressed
version of the spinor helicity formalism and use it to obtain simple
formulae for the flip amplitude with pure external gluon polarizations.
We also give in-depth examples for gauge group SU(2), and evaluate both
the high- and low-energy limits. Throughout, we compare and contrast
with the calculation of photon helicity flip in strong-field QED.}

\maketitle

\section{Introduction}

Background fields play an important role in many physical scenarios, ranging from cosmology and astrophysics to pair production in heavy ion collisions. In the context of quantum electrodynamics (QED), the study of processes in laser fields aims to probe beyond-Standard Model phenomena~\cite{Jaeckel:2010ni,Dobrich:2010hi,Redondo:2010dp} as well as non-perturbative physics within the Standard Model~\cite{Dunne:2008kc}.

The case of strong background fields, which cannot themselves be treated in perturbation theory, is of particular interest both theoretically and experimentally. The theoretical framework for such problems is background perturbation theory~\cite{DeWitt:1967ub,tHooft:1975uxh,Boulware:1980av,Abbott:1981ke}: the strong background is treated \emph{exactly}, with particle scattering occurring perturbatively around the (classical) background. (In QED this is referred to as the Furry expansion~\cite{Furry:1951zz}.) The functional utility of such methods relies, of course, on being able to perform the perturbative calculations without making approximations to the background. In this context simple or highly symmetric backgrounds are natural to consider, and one common choice in QED is to take a plane wave background.  Crucially, the electron propagator in a plane wave is known exactly (the `Volkov propagator'), which enables scattering calculations to be performed explicitly~\cite{Wolkow:1935zz}, see~\cite{Seipt:2017ckc} for a recent review of methods. A plane wave, viewed as a coherent superposition of photons, is commonly employed as a model of intense laser fields; see~\cite{RitusRev,DiPiazza:2011tq,King:2015tba,Seipt:2017ckc} for reviews.

Here we will consider scattering processes in Yang-Mills theory on a plane wave background, the Feynman rules for which were recently determined in~\cite{Adamo:2018mpq}, and which will be extended to QCD here. Yang-Mills plane waves are valued in the Cartan of the gauge group and so are effectively abelian~\cite{Basler:1984hu,Adamo:2017nia}, but perturbative physics around them is fully non-abelian and quite different from that of QED. For instance, \emph{all} fields -- including the gluon -- are charged with respect to the background.

There are several motivations for studying such a system. Firstly, there are few explicit calculations for observables in non-abelian gauge theory in the presence of background fields, especially when the background is treated exactly; this paper demonstrates that such calculations are tractable and lead to concrete results. Second, we believe that what follows constitutes the first calculation of non-abelian gauge theory scattering amplitudes in a background plane wave \textit{beyond} tree-level. Third, the calculation will demonstrate the potential power of modern amplitudes methods in the context of background field calculations for both abelian and non-abelian gauge theories.  We show that the \emph{spinor helicity formalism}~\cite{Srednicki:2007qs,Elvang:2013cua,Dixon:2013uaa,Cheung:2017pzi}, an important tool in the modern study of scattering amplitudes which trivializes on-shell four-dimensional kinematics, generalizes naturally to plane wave backgrounds. Finally, we will be able to compare and contrast non-abelian results with corresponding QED results, offering insights into both. To see this concretely, of course, we need to pick a process to study.

In this paper, we consider the probability for a probe gluon to change helicity upon passing through a strong background gauge field, in both pure Yang-Mills theory with gauge group SU$(N)$, and quantum chromodynamics (QCD) with any number of fundamental flavours. This is the \emph{non-abelian} version of photon helicity flip in QED: a probe photon passing through a strong electromagnetic field has a non-zero probability to change helicity. At leading order, this is a one-loop effect governed by the electron loop diagram in the background. The photon helicity flip amplitude (encoded in the polarization tensor) was calculated long ago for both constant background fields~\cite{Toll:1952rq,Narozhny:1969,Ritus:1972ky,Shore:2007um} and general plane wave backgrounds~\cite{Becker:1974en,Baier:1975ff}. Helicity flip is the process which underpins `vacuum birefringence'~\cite{Toll:1952rq}, a detection target for current optical and X-ray laser experiments; see~\cite{King:2015tba} for a recent review.

The leading contribution to gluon helicity flip is at one-loop on the plane wave background, with three (four) diagrams potentially contributing in pure Yang-Mills (in QCD). While experimental applications for gluon helicity flip are not as immediately obvious as they are for photon helicity flip (due to asymptotic freedom and the lack of control over gluon polarizations), we can nevertheless speculate. In particular, gluon helicity flip may occur for QCD processes in the vicinity of colliding nuclei, which are described using an effective theory known as the Colour Glass Condensate (CGC)~\cite{Iancu:2002xk,Iancu:2003xm,Gelis:2010nm,Kovchegov:2012mbw,Blaizot:2016qgz}. The CGC is composed of high-density, coherent gluonic matter, so it can be modelled by classical colour fields which are strong, and therefore must be treated exactly. The Yang-Mills plane wave backgrounds considered here are in the same class of classical gauge fields which arise in the context of the CGC.

\medskip

This paper is organized as follows: Section~\ref{Background} reviews gauge theory plane waves, and gives the Feynman rules for Yang-Mills and QCD in plane wave backgrounds. In section~\ref{GHF1}, we calculate the one-loop gluon helicity flip amplitude in both Yang-Mills, with gauge group SU$(N)$, and QCD, with any number of fundamental flavours, and for gluons with generic polarizations.  Section~\ref{SHF} demonstrates that the spinor helicity formalism can be used to describe the on-shell kinematics of the probe gluon in the plane-wave background. This enables us to obtain compact expressions for the gluon helicity flip amplitude for pure polarization states: in particular we evaluate the negative-to-positive helicity flip amplitude in section~\ref{GHF2}. We compare throughout with the analogous results for photon helicity flip in QED~\cite{Dinu:2013gaa}. In section~\ref{Exam} we study our formulae using some illustrative examples; we evaluate our expressions in detail for gauge group SU$(2)$, as well as giving the high and low energy limits for arbitrary numbers of colours and flavours. Section~\ref{Conc} concludes with a discussion of future directions as well as potential applications in the context of the CGC.

%%%%%%%%%%%%%%%%%%%%%%%%%%
%%%%%%%%%%%%%%%%%%%%%%%%%%   

\section{Yang-Mills and QCD in a plane wave background}
\label{Background}

The interactions of a single probe gluon with a large number of coherently polarized gluons is modelled by representing the large coherent superposition as a background \emph{plane wave} gauge field. The interactions of the probe with the background and other particles are captured by studying perturbative gauge theory around the background. The Feynman rules associated with non-abelian gauge theory in a plane wave background were recently derived in~\cite{Adamo:2018mpq}; in this section we review the basic structures of perturbative Yang-Mills theory and QCD in a plane wave background, building the toolbox necessary to compute the helicity flip amplitude.  

Recall that the background field approach to perturbative QFT describes Yang-Mills or QCD in terms of a fixed background gauge field $A$ and fluctuating gauge field $\cA$ which is integrated over in the path integral~\cite{DeWitt:1967ub,tHooft:1975uxh,Boulware:1980av,Abbott:1981ke}. The resulting kinetic and interaction terms in the Lagrangian for pure Yang-Mills theory are:
\be\label{blag1}
\mathcal{L}^{\mathrm{YM}}_{\mathrm{kin}}= -\frac{1}{g^2}\,\tr\!\left(D_{[\mu}\cA_{\nu]}\,D^{[\mu}\cA^{\nu]}+\frac{1}{2}F_{\mu\nu}\,[\cA_{\mu},\cA_{\nu}]+\frac{1}{2} (D^{\mu}\cA_{\mu})^2-\frac{1}{4}\bar{c}\,D_{\mu}D^{\mu}\,c\right)\,,
\ee
\be\label{blag2}
\mathcal{L}^{\mathrm{YM}}_{\mathrm{int}}=-\frac{1}{4\,g^2}\tr\!\left(4\,[\cA_{\mu},\,\cA_{\nu}]\,D^{[\mu}\cA^{\nu]}+[\cA_{\mu},\,\cA_{\nu}]\,[\cA^{\mu},\,\cA^{\nu}]-\bar{c}\,D^{\mu}[\cA_{\mu},c]\right)\,,
\ee
where $g$ is the Yang-Mills coupling, $D_{\mu}=\partial_{\mu}-\im [A_{\mu},\cdot]$ is the covariant derivative with respect to the background field, and $F_{\mu\nu}$ is the background field strength. The fermionic ghosts $\{c,\bar{c}\}$ appear as a result of fixing Feynman-'t Hooft gauge for the fluctuating gauge field: $D^{\mu}\cA_{\mu}=0$.  Including quarks of mass $m$ valued in the fundamental representation of the gauge group leads to the additional contributions
\be\label{blag3}
\mathcal{L}^{\mathrm{quark}}_{\mathrm{kin}}= \tr\!\left(\bar{\psi}\,(\im\,\slashed{\partial}+\slashed{A}-m)\psi\right)\,, \qquad \mathcal{L}^{\mathrm{quark}}_{\mathrm{int}}=\tr\!\left(\bar{\psi}\,\slashed{\cA}\,\psi\right)\,.
\ee
The Lagrangians \eqref{blag1} -- \eqref{blag3} define the Feynman rules for Yang-Mills and QCD perturbatively around the background $A$, as we now review for the case of a background plane wave. Throughout, we assume that the gauge group is SU$(N)$; generalizations to other gauge groups are straightforward. 

Note that the background field appears in the kinetic terms for both the gluons and quarks, hence the propagators of the theory are non-trivially dressed by the background. These propagators can be constructed exactly (in particular, without resorting to perturbation theory) when the background is sufficiently simple, or highly symmetric, as is the case for plane waves.

%%%%%%%%%%%%%%%%%%%%%%%%%%

\subsection{Yang-Mills plane waves}

A plane wave gauge field is a highly-symmetric solution of the vacuum Yang-Mills equations which can be viewed quantum mechanically as a coherent state (i.e., a coherent superposition of gluons). In $d$-dimensional Minkowski space, such a gauge field has a $(2d-3)$-dimensional symmetry algebra, isomorphic to a Heisenberg algebra with center given by a covariantly constant symmetry generator; in $d=4$ this corresponds to the Carroll group in 2+1 dimensions with broken rotations~\cite{Carroll,Duval:2017els}. The covariantly constant symmetry is associated to a choice of null direction ($n^{\mu}$, $n^2=0$), which defines the propagation direction of the plane wave. Existence of the Heisenberg symmetry algebra forces the gauge field to be valued in the Cartan of the gauge group~\cite{Adamo:2017nia}.\footnote{While fully non-abelian plane wave solutions have been proposed for Yang-Mills theory in 4-dimensions~\cite{Coleman:1977ps,Kovacs:1978vb,Lo:1979vq}, these do not have the Heisenberg symmetry algebra which seems to be the most natural defining feature of a plane wave solution~\cite{Trautman:1980bj,Basler:1984hu}. Furthermore, it is unclear how to generalise the criteria defining these non-abelian solutions to arbitrary dimension, whereas the symmetry definition is essentially dimension-independent. This will be discussed further elsewhere.} While our primary focus will be $d=4$ physics, we review the general features of Yang-Mills plane waves for $d$ arbitrary.

Let the Minkowski metric in light cone coordinates $(x^{\LCp}, x^{\LCperp}, x^{\LCm})$ be given by
\be\label{Minkmet}
\ud s^2 = 2\, \ud x^{\LCp}\,\ud x^{\LCm} - \left(\ud x^{\LCperp}\right)^2\,.
\ee
In these coordinates it is convenient to choose $n^{\mu}=\delta^{\mu}_{\LCm}$, whence $n\cdot x = x^\LCm$ and the gauge potential for a Yang-Mills plane wave can be written as
\be\label{gaugefield1}
A = x^\LCperp \dot{a}_\LCperp(x^\LCm)\, \ud x^\LCm\,,
\ee
with the $d-2$ Cartan-valued $a_{\LCperp}$ being free functions of $x^{\LCm}$. Note that the gauge field \eqref{gaugefield1} is functionally equivalent to the well-known electromagnetic plane wave (cf., \cite{Wolkow:1935zz,Schwinger:1951nm,Heinzl:2017blq}); it differs only by being valued in a more general Cartan subgroup than U$(1)$.

Choosing this gauge to express the potential has the benefit that the field strength $F$ is encoded algebraically in $A$; indeed
\be\label{fs}
F=\dot{a}_{\LCperp}(x^{\LCm})\,\d x^{\LCperp}\wedge\d x^{\LCm}\,.
\ee
In any physically reasonable scenario, it is clear that the gauge field strength should have finite extent in time. This means that $\dot{a}_{\LCperp}(x^{\LCm})$ is compactly supported in $x^{\LCm}$, ensuring that there are asymptotic `in' and `out' regions necessary to define an S-matrix perturbatively around the plane wave background. Indeed, it can be shown that the S-matrix for gluon perturbations on this plane wave background is well-defined in the sense of unitary evolution~\cite{Adamo:2017nia}. We therefore restrict our attention to such `sandwich' plane wave backgrounds in Yang-Mills theory from now on. Note that, in common with both the electromagnetic~\cite{Schwinger:1951nm} and gravitational~\cite{Gibbons:1975jb,Deser1975} cases, there is no spontaneous (Schwinger) particle creation in plane wave backgrounds.

%%%%%%%%%%%%%%%%%%%%%%%%%%%%%%

\subsection{External legs in scattering amplitudes}

External particles in a scattering process on the plane wave are initially free fields which propagate from the in-region of space-time (where $\dot{a}_{\LCperp}=0$), across the non-trivial plane wave background, and then to the asymptotic future of the out-region (where $\dot{a}_{\LCperp}=0$ once again). The LSZ reduction states that these external states -- corresponding to the external legs of any Feynman diagrams on the plane wave -- are described by solutions to the free equations on the background. Since the asymptotic states are defined in regions where the gauge field is trivial, it is clear that they are uniquely specified by the same quantum numbers as in perturbation theory around a globally trivial background, namely an on-shell momentum, a polarization, and a colour or flavour vector. 

\subsubsection*{\textit{Gluons}}

On a flat background, an on-shell gluon is characterised by momentum $k_{\mu}$ and polarization $\varepsilon_{\mu}$ obeying $k^2=0=k\cdot\varepsilon$, as well as a generator of the gauge group $\mathsf{T}^{\sa}$, where $\mathsf{a}=1,\ldots N^2-1$ for SU$(N)$. The additional light cone gauge constraint $n\cdot\varepsilon=0$ can also be imposed. The resulting gluon is then described by the usual asymptotic wavefunction $\mathsf{T}^{\sa}\varepsilon_{\mu} \e^{\im k\cdot x}$. In the presence of a plane wave background, the flat background wavefunction is modified to
\be\label{gluonwf}
	\mathsf{T}^{\sa}\,\mathcal{E}_\mu(x^{\LCm})\,  \exp \left[\im\,\phi_k(x)\right] \,,
\ee
in which the function $\phi_k(x)$ and vector $\mathcal{E}_\mu(x^{\LCm})$ reduce to $k\cdot x$ and $\varepsilon_\mu$ respectively when the background is turned off. Here and throughout $k_\mu$ and $\varepsilon_\mu$ should be thought of as the momentum and polarization of the gluon \textit{before} it enters the plane wave~\cite{Dinu:2012tj,Ilderton:2012qe}. The quantities $\phi_k$ and $\mathcal{E}_\mu$ then encode the effect on $k_\mu$ and $\varepsilon_\mu$ of interactions with the gluonic background, as follows. The exponent in \eqref{gluonwf} defines a dressed momentum $K_\mu(x^\LCm)$ via
\begin{equation*}
 -\im\,\e^{-\im\phi_k}D_{\mu}\,\e^{\im\phi_k}  = K_\mu(x^\LCm)\,,
\end{equation*} 
which is on-shell, $K^2(x^\LCm)=0$. This dressed momentum is that of a classical particle moving under the Lorentz force due to the background plane wave. The vector $\mathcal{E}_\mu(x)$ is the dressed polarization which is transverse to the dressed Lorentz momentum, so $ K_\mu \mathcal{E}^\mu = 0$. Hence~(\ref{gluonwf}) is a one-particle wavefunction for the gluon.

Explicitly, $\phi_k(x)$ is a solution to the gauge-covariant Hamilton-Jacobi equations on the plane wave background, given by~\cite{Adamo:2017nia}
\be\label{HamJac}
	\phi_k(x) = k_{\LCp}\,x^{\LCp}+\left(k_{\LCperp}+ea_{\LCperp}(x^\LCm)\right)\,x^{\LCperp}+\frac{1}{2\,k_{+}}\,\int^{x^\LCm}\!\!\left(k_{\LCperp} +ea_{\LCperp}(\alpha)\right)^2\, \ud \alpha \,.
\ee
Here, $e$ represents the \emph{colour charge} of the initial gluon with respect to the background, taking values in a root space. In particular, if $\{\mathsf{T}^{\mathsf{i}}\}$ are the $N-1$ generators of the Cartan of SU$(N)$, then the charge is $(e^{\sa})^{\mathsf{i}\mathsf{b}}:=[\mathsf{T}^{\mathsf{i}},\mathsf{T}^{\mathsf{b}}]^{\mathsf{a}}$. The explicit colour indices are suppressed in \eqref{HamJac}, since the contractions are obvious. This notation enables us to compactly represent commutators between the Cartan-valued background and gluon perturbations with arbitrary colour, since $[a_{\LCperp}, \mathsf{T}^{\sa}]=e a_{\LCperp}\,\mathsf{T}^{\sa}$. (Note that we will sometimes abuse notation and separate the charge from the field, writing for example $ea_{\LCperp} \cdot e a_{\LCperp} = e^2 a_{\LCperp}^2$.) Using the explicit form of $\phi_k$ it follows that
\be\begin{split}\label{K-DEF}
	K_\mu(x^\LCm) =   k_\mu + e a_\mu - \frac{1}{2k_\LCp} n_\mu \left(2ea_\sigma k^\sigma + e^2 a_\sigma a^\sigma\right) \,,
\end{split}
\ee
for $a_{\mu}=\delta_{\mu}^{\LCperp}a_{\LCperp}$ and $n_{\mu}=\delta_{\mu}^{\LCm}$. It can be checked directly that the dressed momentum obeys $K^2=0$. $K_{\mu}$ is related to the initial momentum $k_{\mu}$ by
\be\begin{split}\label{K-DEF2}
	K_\mu(x^\LCm) &= \bigg(\eta_{\mu\nu}  + \frac{1}{k_{\LCp}} \big( e a_\mu\, n_\nu - n_\mu\, ea_\nu \big) - \frac{1}{2\,k_{\LCp}^2} e^2a^\sigma a_\sigma\, n_\mu n_\nu \bigg) k^\nu \\
		& =: K_{\mu\nu}(x^{\LCm})\, k^\nu \;.
\end{split}
\ee
It is easily verified that $K_{\mu\nu}(x^{\LCm}) = \exp\big( [ea,n]/k_{\LCp}\big)_{\mu\nu}$ is a Lorentz boost, hence why $K_\mu$ is on-shell if $k_\mu$ is. Similarly, the dressed polarization vector $\mathcal{E}_{\mu}$ appearing in \eqref{gluonwf} is the boosted initial polarization:
\be\label{EpolDef}
	\mathcal{E}_\mu(x^{\LCm}) = K_{\mu\nu}(x^{\LCm})\, \varepsilon^\nu = \bigg( \eta_{\mu\nu}  -\frac{1}{k_{\LCp}} n_\mu\, ea_\nu \bigg) \varepsilon^\nu \;,
\ee
again making it clear why $K(x^\LCm) \cdot \mathcal{E}(x^\LCm)=0$, and also that $\mathcal{E}_\mu$, like $\varepsilon_\mu$, obeys the lightcone gauge condition $n \cdot \mathcal{E} = 0$. It is a straightforward exercise to verify that \eqref{gluonwf} obeys the linearised Yang-Mills equation on the plane wave background with these definitions.

\subsubsection*{\textit{Quarks}}

Similarly, an asymptotic quark state is characterised by an on-shell momentum $k_\mu$, a Dirac spinor $u_k$ (obeying $k^2=m^2$ and $\slashed{k}u_k=-mu_k$) and a vector $\underline{\mathsf{t}}$ of the fundamental representation of SU$(N)$. In the presence of the plane wave background, both the momentum and the Dirac spinor become dressed. For fundamental matter, such as the quarks, the colour charge $e$ appearing for gluon wavefunctions is replaced by a weight $\mu$ of the fundamental representation. Since the background gauge field is valued in the Cartan, we have 
\begin{equation*}
a_{\LCperp}\,\underline{\mathsf{t}}=a_{\LCperp}^{\mathsf{i}}\,\mathsf{T}^{\mathsf{i}}\underline{\mathsf{t}}=a^{\mathsf{i}}_{\LCperp}\,\mu^\mathsf{i}\,\underline{\mathsf{t}}\equiv\mu a_{\LCperp}\,\underline{\mathsf{t}}\,,
\end{equation*}
where contractions in the Cartan are again implicit and the dependence of the fundamental weight $\mu$ on $\underline{\mathsf{t}}$ has been suppressed. With this the quark wavefunction becomes
\be\label{qwf}
	\underline{\mathsf{t}}\,\bigg(\mathbb{I}-\frac{\slashed{n}\,{\mu\slashed{a}}}{2\,k_{\LCp}} \bigg) \cdot u_k\,\exp\left[\im\,\tilde{\phi}_k(x)\right]\,,
\ee
in which $\tilde{\phi_k}$ reduces to $k\cdot x$ in a flat background, and where the spinor matrix can be determined by solving the Dirac equation (or demanding its consistency upon contraction with the adjoint Dirac operator). The function $\tilde{\phi}_k$ is given by
\be\label{tphi}
\tilde{\phi}_k(x) = k_{\LCp}\,x^{\LCp}+\left(k_{\LCperp}+\mu\,a_{\LCperp}(x^\LCm)\right)\,x^{\LCperp}+\frac{1}{2\,k_{+}}\,\int^{x^\LCm}\!\!\left[m^2+\left(k_{\LCperp} +\mu\,a_{\LCperp}(\alpha)\right)^2\right]\, \ud \alpha \,\,,
\ee
which differs from the exponent in the gluon wavefunction \eqref{HamJac} only through the mass term.

Just as there is a dressed momentum associated with the gluon state in a plane wave background, the dressed momentum associated with the quark state \eqref{qwf} is
\be\label{qmom}
	\tilde{K}_{\mu}(x^{\LCm})= k_\mu + \mu\, a_\mu - \frac{n_{\mu}}{2k_\LCp} \left(2\mu\,a_\sigma k^\sigma + \mu^2\, a_\sigma a^\sigma\right)\,,
\ee
where the tilde on $\tilde{K}_{\mu}$ is to distinguish it as a \emph{massive} dressed momentum, obeying $\tilde{K}^{2}=m^2$. The spinor structure is also explicitly dressed in \eqref{qwf}, and has a simple physical interpretation; it is the Lorentz-boosted free spinor, and is equal to the free spinor for momentum ${\tilde K}_\mu$ rather than $k_\mu$, so 
\be
	\bigg(\mathbb{I}-\frac{\slashed{n}\,{\mu\slashed{a}}}{2\,k_{\LCp}} \bigg) \cdot u_k \equiv u_{\tilde{K}(x^\LCm)} \;.
\ee
In the case that the gauge group is U$(1)$, the fundamental weights obey $\mu\rightarrow 1$ and these quark solutions reduce to the well-known Volkov solutions of QED for an electron in a background plane wave~\cite{Wolkow:1935zz}. For a recent review of their properties see~\cite{Seipt:2017ckc}, for applications to BSM physics~\cite{Heinzl:2009zd,VillalbaChavez:2012bb,Dillon:2018ypt,King:2018qbq}, for neutrino physics~\cite{Meuren:2015iha}, and for helicity flip, which underlies vacuum birefringence~\cite{Heinzl:2006xc,Karbstein:2015xra,Schlenvoigt2016}, see~\cite{Dinu:2013gaa,King:2015tba}.

%%%%%%%%%%%%%%%%%%%%%%%%%%%%%%%%

\subsection{Propagators and vertices}

Propagators for the gluon and quark fields appearing in Yang-Mills and QCD in a plane wave background can be obtained by taking a sum over the linearised states defined above. The Feynman propagator for the gluon in Feynman-'t Hooft gauge was obtained in~\cite{Adamo:2018mpq}:
\be\label{gluonp}
	\scG^{\sa\mathsf{b}}_{\mu\nu}(x,y)= -\im\, \delta^{\sa\mathsf{b}}\,\oint\! \frac{\d^{d}k}{(2\pi)^d}\,\frac{\D^{k}_{\mu\nu}(x^{\LCm},y^{\LCm})}{k^{2}+\im\,\varepsilon}\,\exp\left[\im\phi_k(x)-\im\phi_{k}(y)\right]\,,
\ee
where the integral is over the $d$ parameters $\{k_{\LCp},k_{\LCperp},k_{\LCm}\}$ of an off-shell momentum, taken over the usual Feynman contour in the $k_{\LCm}$ plane. The non-trivial tensor structure of the propagator, $\D_{\mu\nu}^{k}(x^{\LCm},y^{\LCm})$, is given by
\be\label{tstruct}
\D^{k}_{\mu\nu}(x^{\LCm},y^{\LCm}):=K_{\mu\sigma}(x^{\LCm})\,K_{\nu}{}^{\sigma}(y^{\LCm})\,,
\ee
where $K_{\mu\nu}(x^{\LCm})$ is defined in \eqref{K-DEF2}, continued off-shell.

For the scalar (Grassmann) ghosts, the propagator is obtained simply by dropping the tensor structure in the gluon propagator, so~\cite{Adamo:2018mpq}
\be\label{ghostp}
	\scG^{\sa\mathsf{b}}(x,y) = -\im\,\delta^{\sa\mathsf{b}}\,\oint\! \frac{\d^{d}k}{(2\pi)^d}\,\frac{1}{k^{2}+\im\,\varepsilon}\,\exp\left[\im\phi_k(x)-\im\phi_{k}(y)\right]\,,
\ee
It is a straightforward exercise to demonstrate that \eqref{gluonp} and \eqref{ghostp} are Green's functions for the appropriate differential operators.

For quarks, the propagator is obtained by summing over the linearised states \eqref{qwf}:
\be\label{quarkp}
	\scG^{ij}(x,y) =  \im\, \delta^{A}_{B}\, \delta^{ij}\oint\frac{\ud^d k}{(2\pi)^d} \, \frac{V^{k}(x^{\LCm},y^{\LCm})}{k^2-m^2+\im\,\varepsilon}\, \exp\left[\im\tilde{\phi}_k(x) - \im\tilde{\phi}_k(y)\right]\,,
\ee
in which the Kronecker delta $\delta^{A}_{B}$ is over the fundamental/anti-fundamental representation of the gauge group, $\delta^{ij}$ is over flavour indices and $V^k$ contains the the spin structure,
\be\label{VolkDef}
	 V^k(x^{\LCm},y^{\LCm}) := \left(\mathbb{I} - \mu\,\frac{\slashed{n}\, {\slashed{a}}(x^{\LCm})}{2\,k_{\LCp}}\right) \left(-\slashed{k}+m\right) \left(\mathbb{I} + \mu\,\frac{\slashed{n}{\slashed{a}}(y^{\LCm})}{2\,k_{\LCp}}\right) \,.
\ee
This matches the standard literature representation of $V^k$ in QED, but it proves cumbersome in calculations due to the number of different gamma-matrix structures appearing. However, it can be rewritten in a more useful and revealing form as
\be\label{VolkNew}
\begin{split}
	 V^k(x^{\LCm},y^{\LCm}) &= -\frac{1}{2\,k_{\LCp}} (\slashed{\tilde K}(x^{\LCm})-m)\, \slashed{n}\,(\slashed{\tilde K}(y^{\LCm}) -m ) + \slashed{n}\,\frac{k^2-m^2}{2\,k_{\LCp}} \\
	 	&:= \vec{V}^{K}(x^{\LCm},y^{\LCm}) + I^k\,,
\end{split}
\ee
which we do not believe has appeared in the literature before. One immediate advantage of the representation \eqref{VolkNew} is that it expresses the propagator in terms of the dressed momenta. Additionally, the contribution of $I^{k}$ to the propagator \eqref{quarkp} has no momentum pole, so upon performing the momentum integrals this term becomes proportional to $\delta(x^\LCm - y^\LCm)$. This is the `instantaneous propagator' of lightfront zero modes (cf., \cite{Mustaki:1990im,Schoonderwoerd:1998qj,Srivastava:2000cf,Mantovani:2016uxq}), which is manifest in the representation \eqref{VolkNew}.

\medskip

\begin{figure}[t!]
\centering
\includegraphics[scale=0.75]{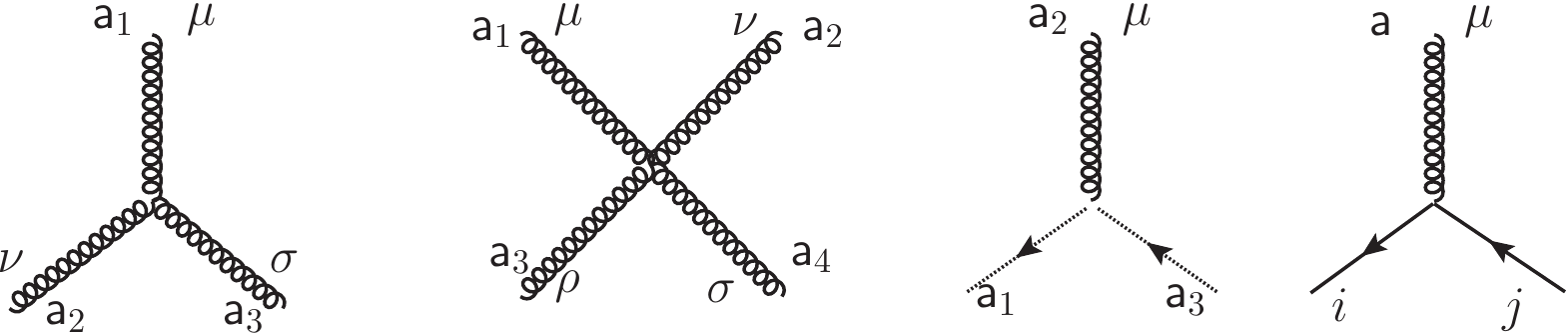}
\caption{\label{FIG:VERTS} From left to right, the gluon, ghost and fermion vertices in Yang-Mills and QCD.}
\end{figure}

The list of Feynman rules for perturbative Yang-Mills and QCD on the plane wave background is completed by specifying the vertices of the theory. These are easily read off from the interacting contributions to the background field Lagrangian \eqref{blag2}, \eqref{blag3}. The gluon 3-point and 4-point interaction vertices are
\be\label{gl3pv}
g\,f^{\sa_{1}\sa_{2}\sa_{3}}\,\int \d^{d}x\,\left(\eta^{\mu\nu}\,(D_{1}-D_2)^{\sigma}+\eta^{\nu\sigma}\,(D_{2}-D_{3})^{\mu}+\eta^{\sigma\mu}\,(D_{3}-D_{1})^{\nu}\right)\,,
\ee
\begin{multline}\label{4pv}
g^{2}\,\int \d^{d}x\,\left[f^{\sa_{1}\sa_{2}\mathsf{b}}f^{\sa_{3}\sa_{4}\mathsf{b}}\,(\eta^{\mu\rho}\eta^{\nu\sigma}-\eta^{\mu\sigma}\eta^{\nu\rho}) \right. \\
\left. + f^{\sa_{1}\sa_{3}\mathsf{b}}f^{\sa_{2}\sa_{4}\mathsf{b}}\,(\eta^{\mu\nu}\eta^{\rho\sigma}-\eta^{\mu\sigma}\eta^{\nu\rho})+f^{\sa_{1}\sa_{4}\mathsf{b}}f^{\sa_{2}\sa_{3}\mathsf{b}}\,(\eta^{\mu\nu}\eta^{\rho\sigma}-\eta^{\mu\rho}\eta^{\nu\sigma})\right]\,.
\end{multline}
Here, $f^{\sa_1\sa_2\sa_3}$ are the structure constants of SU$(N)$ and the covariant derivative $D_{i}$ is understood to act on leg $i$ of the relevant vertex. See Figure~\ref{FIG:VERTS} for the conventions regarding assignment of indices. Furthermore, there is an implicit conservation of charge with respect to the background gauge field at each vertex. 

The interaction vertex between ghosts and a gluon is 
\be\label{ghostv}
g\,f^{\sa_{1}\sa_{2}\sa_3}\,\int \d^{d} x\,D^{\mu}_{1}\,,
\ee 
while the quark-gluon vertex is
\be\label{quarkv}
g\,\delta^{ij}\, \mathsf{T}^{\sa}\,\gamma^{\mu} \int\d^{d}x\,\,.
\ee
It is easy to see that in the flat background limit, all of these expressions can be evaluated on momentum eigenstates to return the familiar momentum space Feynman rules of QCD.

%%%%%%%%%%%%%%%%%%%%%%%%%%%%%%%%%%%%%%%%%%%%%%%%%%%%%%%
%%%%%%%%%%%%%%%%%%%%%%%%%%%%%%%%%%%%%%%%%%%%%%%%%%%%%%%

\section{Gluon helicity flip}
\label{GHF1}

The probability for a probe gluon to flip helicity after traversing the plane wave background is encoded in the $1\rightarrow 1$ gluon scattering amplitude. In all amplitudes on plane wave backgrounds one has conservation of three momenta, $k_\LCp$ and $k_\LCperp$, following from the invariance of the plane wave under translations in the $x^\LCp$ and $x^\LCm$ directions. However, momentum $k_\LCm$ is not conserved, due to the arbitrary dependence of the plane wave background on $x^{\LCm}$. It follows that the $1\to1$ gluon scattering amplitude will take the form
\be\label{SFI}
	S_{fi} = (2\pi)^{3}\, \delta^{3}_{\LCperp,\LCp}(k+k')\, M(k) \;,
\ee
where the incoming gluon has (un-dressed) momentum $k_{\mu}$ and polarization $\varepsilon_\mu$, before entering the plane wave, while the outgoing gluon carries momentum $k'_{\mu}$ and polarization~$\varepsilon'_\mu$ after leaving the wave. As can be seen from \eqref{SFI}, scattering without emission in a plane wave background is forward. It follows that the tree-level contribution to the helicity flip amplitude is zero, since the $1\rightarrow 1$ tree-level gluon amplitude on the background is proportional to $\varepsilon\cdot\varepsilon'$. Therefore, the leading contribution to gluon helicity flip is a one-loop effect.

\begin{figure}[t!]
\includegraphics[width=0.51\textwidth]{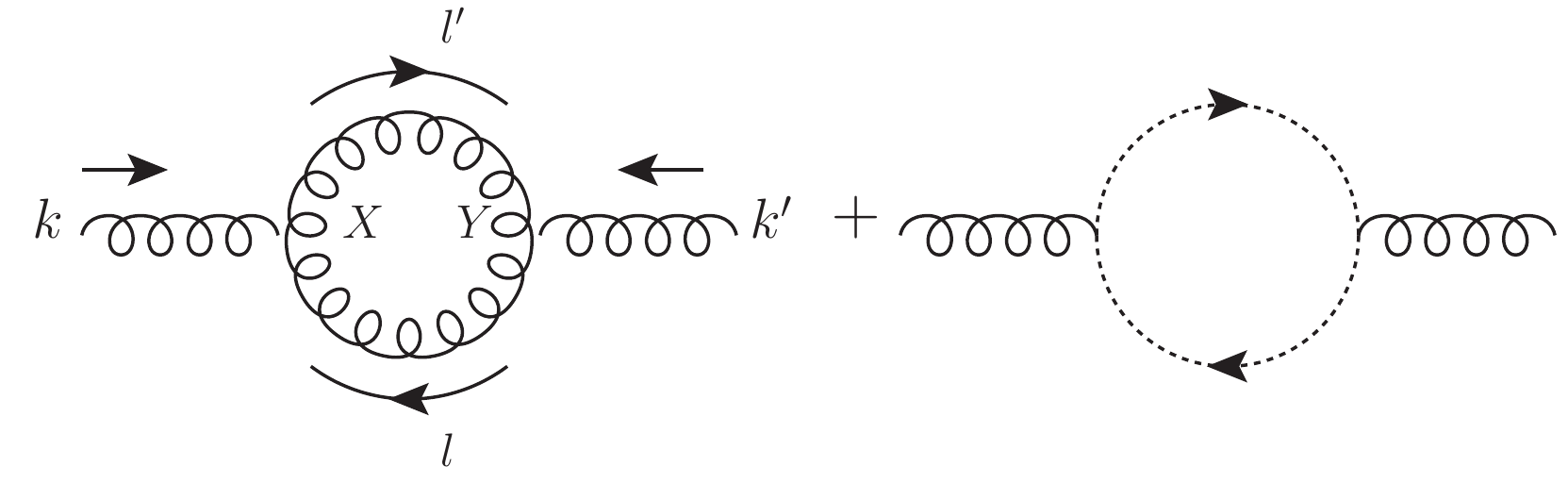}\raisebox{9pt}{\includegraphics[width=0.43\textwidth]{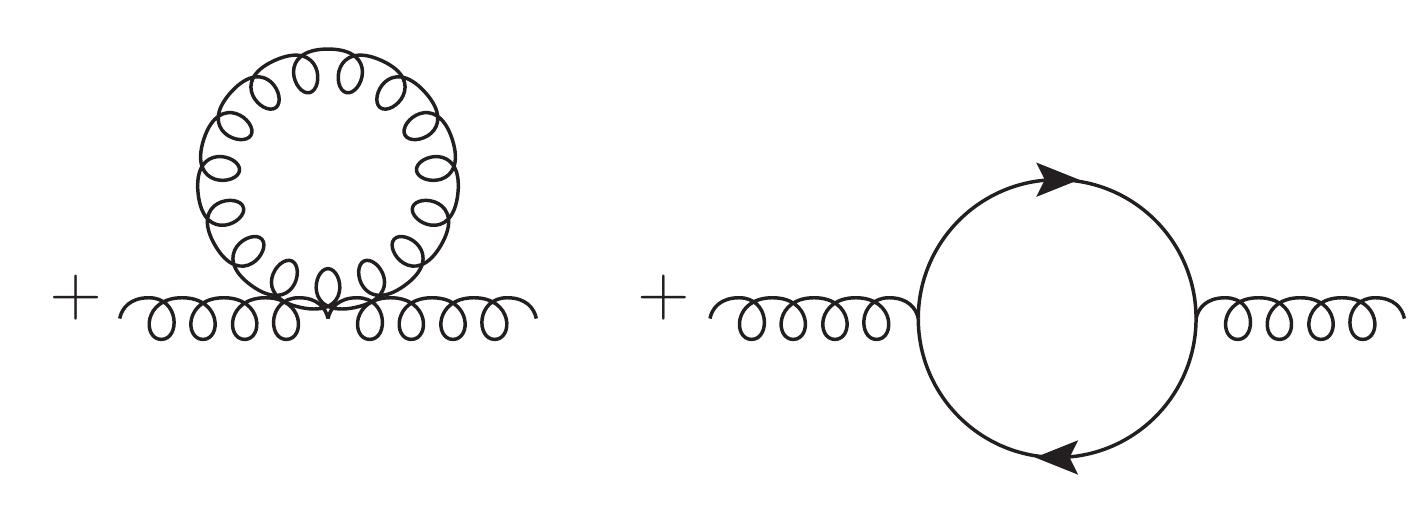}}
\caption{\label{FIG:LOOPS} Diagrams contributing to helicity flip at one loop: the gluon and ghost loops, the gluon tadpole and the quark loop. Momentum assignments are the same in all diagrams. All particles are charged with respect to the background, so that the corresponding propagators and external legs are to be understood as dressed.}
\end{figure}

The contributing Feynman diagrams are shown in Fig.~\ref{FIG:LOOPS}; the quark loop may be dropped to obtain the amplitude for pure Yang-Mills theory.  Including a normalised wavepacket for the initial state, one can show that for the amplitude \eqref{SFI} the total probability of helicity flip is~\cite{Ilderton:2012qe}
\be
	\mathbb{P}_\text{flip} = \bigg|\frac{M(k)}{2k_\LCp}\bigg|^2 \;.
\ee
Thus our focus will be on the nontrivial part of the scattering amplitude, $M(k)$, which implicitly contains dependence on the polarization vectors, from here on obeying $\varepsilon\cdot\varepsilon'=0$ for helicity flip.

%   

%indeed, this is a consequence of the `no particle production' statement for plane wave backgrounds which ensures that the S-matrix is well-defined~\cite{Schwinger:1951nm,Adamo:2017nia}. \Anton{[Hmmm, think about that]} 

%%%%%%%%%%%%%%%%%%%%%%%%%%

\subsection{Diagram contributions}

Using the Feynman rules for perturbative Yang-Mills and QCD in a plane wave background defined in section \ref{Background}, we now calculate the contributions to the amplitude $M(k)$ from each of the diagrams in Fig.~\ref{FIG:LOOPS}. These calculations are performed in general dimension $d$ to enable dimensional regularization of potentially divergent integrals prior to fixing $d=4$ at a later stage.

\subsubsection{Ghost loop}

The ghost loop has the simplest tensor structure (as the ghosts running in the loop are scalars), making it the natural starting point. Furthermore, the underlying method of calculation described here applies analogously to the gluon and quark loop diagrams.

The covariant derivatives in the vertices \eqref{ghostv} bring down factors of the dressed momenta which are immediately contracted with the dressed polarization tensors of the external gluons, giving $\mathcal{E}(x)\cdot L'(x) \mathcal{E}'(y)\cdot L(y)$ in the integrand where $\mathcal{E}_\mu$ and $\mathcal{E}'_\mu$ are the dressed incoming and outgoing polarizations, respectively. Next, the majority of the integrals can be performed. The dependence of the integrand on the transverse ($x^\LCperp$, $y^\LCperp$) and longitudinal coordinates ($x^\LCp$, $y^\LCp$) is trivial, and performing the integrals over these variables yields the overall momentum-conserving delta functions in \eqref{SFI}, as well as a second set of delta functions fixing $l'_\LCperp = l_\LCperp + k_\LCperp$ and $l'_\LCp = l_\LCp + k_\LCp$. The integrals over $l'_{\LCperp}$ and $l'_{\LCp}$ can be performed trivially against these latter delta functions.

We now perform the integrals over $l_\LCm$ and $l'_\LCm$. We use the residue theorem, but because of the dependence of the background on $x^\LCm$, the integrals are performed as in lightfront field theory~\cite{Brodsky:1997de,Heinzl:2000ht}. Recall that in our lightfront coordinates, the mass-shell is $l^2 = 2 l_\LCp l_\LCm - l_\LCperp^2=0$, and consider the $l_\LCm$ integral:
\be\begin{split}\label{L-INT}
	\frac{1}{2\pi \im}\oint\!\ud l_\LCm \frac{\e^{\im\, l_\LCm(x-y)^\LCm}}{l^2+\im\varepsilon} &= \frac{1}{2\pi \im} \oint\! \frac{\ud l_\LCm}{2l_\LCp} \frac{\e^{\im\, l_\LCm(x-y)^\LCm}}{l_\LCm  - \frac{l_\LCperp^2-\im\varepsilon}{2l_\LCp}} \\
	&=\frac{e^{\im\, l_\LCm^{\mathrm{o.s.}}(x-y)^\LCm}}{2l_\LCp} \bigg[\Theta(x^\LCm - y^\LCm)\Theta(-l_\LCp) - \Theta(y^\LCm - x^\LCm)\Theta(l_\LCp)\bigg] \;.
\end{split}
\ee
In the first line we have written out $l^2$ in terms of components to show that the position of the pole switches between the upper and lower half $l_\LCm$ plane depending on the sign of $l_\LCp$. In the second line we have performed the integral, which puts $l_\LCm$ on-shell at $l_\LCperp^2/(2l_\LCp)$. The $\ud l'_\LCm$ integral is performed in exactly the same way, except that $l'_\LCm$ is put on-shell for transverse and longitudinal momenta given by $l+k$, hence $l'_\LCm \to l^{\prime \mathrm{o.s.}}_\LCm := (l+k)_\LCperp^2/2(l+k)_\LCp$, and one obtains instead of \eqref{L-INT}
\be\begin{split}\label{LP-INT}
	\frac{e^{\im\, l^{\prime \mathrm{o.s.}}_\LCm(y-x)^\LCm}}{2(l+k)_\LCp} \bigg[-\Theta(y^\LCm - x^\LCm)\Theta(-l_\LCp-k_\LCp) + \Theta(x^\LCm - y^\LCm)\Theta(l_\LCp+k_\LCp) \bigg] \;.
\end{split}
\ee
Note the exchanged position dependence on $x^\LCm$, $y^{\LCm}$. The product of \eqref{L-INT} and \eqref{LP-INT} appears in the amplitude. Since\footnote{This is true for all particles, except massless particles with momentum aligned with $n^\mu$; these are the infamous zero modes of lightfront field theory~\cite{Brodsky:1997de,Heinzl:2000ht,Heinzl:2003jy}. Such modes propagate parallel to, or \textit{collinear} with, the background plane wave. We discuss this case in Sect.~\ref{SECT:IR}.} $k_\LCp>0$, only the product of terms with $\Theta(x^\LCm-y^\LCm)$ can contribute, and so $l_\LCp$ is constrainted to lie in the range $-k_\LCp< l_\LCp < 0$.

\medskip 

It is useful to take stock at this stage. There are $(d+1)$ integrals remaining. Two, coming from the vertices, are over the lightfront times $x^\LCm$ and $y^\LCm$ on which the background depends. These cannot be performed analytically in general (see section~\ref{Exam} for limits). The remaining $(d-1)$ integrals are over $l_\LCperp$ and $l_\LCm$, the only loop momenta components unconstrained by delta functions. Note also that the integrations have fixed $l'_{\LCp,\LCperp}=(l+k)_{\LCp,\LCperp}$, and $l'_{\LCm}$ has been put on-shell by the contour integral. Thus, the fixed $l'_\mu$ may be written more covariantly as 
\be\label{l-replace-1}
	l'_\mu \to l_\mu + k_\mu - \frac{(l+k)^2}{2\,(l+k)_{\LCp}}\, n_\mu\,, 
\ee
as may be verified using the mass-shell condition. It follows that the dressed momentum~$L'$ is also fixed to
\be\label{l-replace-2}
	L'_\mu \to L_\mu + K_\mu -\frac{(L+K)^2}{2\,(l+k)_{\LCp}}\, n_\mu \;.
\ee
Note that, because $\mathcal{E} \cdot K(x) = n\cdot\mathcal{E}(x) = 0$, the essential tensor structure in the integrand, $\cE(x) \cdot L'(x) \cE'(y)\cdot L(y)$ simplifies to
\be
	\cE\cdot L(x)\,\cE'\cdot L(y) \;,
\ee
abbreviating $\cE(x)\cdot L(x) \equiv \cE\cdot L(x)$, etc.

The integrand also contains exponential terms which we have not yet discussed. At the start, we had in the exponent ($\im$ times)
\be
	\phi_k(x) + \phi_{k'}(y) + \phi_l(x) - \phi_l(y) + \phi_{l'}(y) - \phi_{l'}(x) \;,
\ee
in which the first two terms came from the external legs, and the remaining terms from the loop propagators. This simplifies considerably after the above integrations are performed and charge conservation ($e_l' = e_l + e_k$) is imposed, leaving in the exponent ($\im$ times)
\be\begin{split}\label{i-exponenten}
%	&\int\limits_{y^\LCm}^{x^\LCm}\!\ud s\, \frac{1}{2k_\LCp}(k_\LCperp+eA_\LCperp)^2  + \frac{1}{2l_\LCp}(l_\LCperp+e_lA_\LCperp)^2 - \frac{1}{2(l_\LCp+k_\LCp)}(l_\LCperp+e_lA_\LCperp + k_\LCperp + eA_\LCperp)^2  \\
	%
%	&=\frac{1}{2(l_\LCp+k_\LCp)}\int\limits_{y^\LCm}^{x^\LCm}\!\ud s\, \bigg(1 + \frac{l_\LCp}{k_\LCp}\bigg) K_\LCperp^2 + \bigg(1 + \frac{k_\LCp}{l_\LCp}\bigg) L_\LCperp^2 - (L_\LCperp + K_\LCperp)^2 \\
%	&=\frac{1}{(l_\LCp+k_\LCp)}\int\limits_{y^\LCm}^{x^\LCm}\!\ud s\, \frac{l_\LCp}{2k_\LCp} K_\LCperp^2 + \frac{k_\LCp}{2l_\LCp} L_\LCperp^2 - L_\LCperp K_\LCperp =
 \frac{1}{(l+k)_\LCp}\int\limits_{y^\LCm}^{x^\LCm}\!\ud \alpha\, K(\alpha)\!\cdot\! L(\alpha) \;.
\end{split}
\ee
This is very similar to the exponents which appear at both tree level and one loop for QED processes, with the key difference that in \eqref{i-exponenten} \emph{all} momenta are position dependent and appear under the integral since gluons are charged with respect to the background.

To present the final result, and to better compare with QED results, we change variable from $l_\LCm$ to the lightfront momentum fraction $s:= - l_\LCp/k_\LCp$, so $0<s<1$. In these variables, the full contribution from the ghost loop is:
\be\begin{split}\label{ghost-final-1}
	M_\text{ghost}(k) = -\frac{g^2 C_2(G)\delta^{\mathsf{ab}}}{4 k_\LCp\, (2\pi)^{d-1}} &\int\limits_{-\infty}^{+\infty}\!\ud x^\LCm\!\int\limits^{x^\LCm}_{-\infty}\ud y^\LCm\! 
	\int\limits_0^1\!\frac{\ud s}{s(1-s)} \int\!\ud^{d-2}l_\LCperp
	\, \e^{ \frac{\im}{k_{\LCp}(1-s)}\int\limits_{y^\LCm}^{x^\LCm}\!\ud\alpha\, K\cdot L(\alpha)} \\
	&\times \cE\cdot L(x)\,\cE'\cdot L(y)\,, 
\end{split}
\ee
with the overall factor of $-1$ due to the fermionic statistics of the ghosts, and $C_2(G)$ the quadratic Casimir of the gauge group. Note that the integrals over the transverse loop momentum $l_\LCperp$ are Gaussian and can be performed immediately. The explicit result depends on the chosen external helicity states (in particular, how $l_\LCperp$ appears in $\mathcal{E}\cdot L(x) \mathcal{E}'\cdot L(y)$), so we delay this final integral until later.

\subsubsection{Gluon loop}

For the gluon loop, both the position and loop momentum integrals go through as before. All exponential terms are precisely as for the ghost loop. We can therefore write the gluon contribution in the same way as \eqref{ghost-final-1},
\be\begin{split}\label{gluon-final-1}
	M_\text{gluon}(k) = \frac{g^2 C_2(G)\delta^{\mathsf{ab}}}{4 k_\LCp (2\pi)^{d-1}} &\int\limits_{-\infty}^{+\infty}\!\ud x^\LCm\!\int\limits^{x^\LCm}_{-\infty}\ud y^\LCm\! 
	\int\limits_0^1\!\frac{\ud s}{s(1-s)} \int\!\ud^{d-2}l_\LCperp
	\, \e^{ \frac{\im}{k_{\LCp}(1-s)}\int\limits_{y^\LCm}^{x^\LCm}\!\ud\alpha\, K\cdot L(\alpha)} \\
	&\times \frac12 \mathcal{M}(x^{\LCm},y^{\LCm})  \;,
\end{split}
\ee
where $\frac{1}{2}$ is the symmetry factor for the loop, and the difficult part of the calculation is identifying $\mathcal{M}$ which comes entirely from the tensor structure. This is significantly more complicated than that of the other diagrams, and is given by:
\begin{multline}\label{swamp1}
%\left[\eta^{\mu\rho} (K-L)^{\sigma}+\eta^{\rho\sigma} (L+L')^{\mu}-\eta^{\sigma\mu} (L'+K)^{\rho}\right](x)\, \cE_{\mu}(x)\,\cE'_{\nu}(y) \\
%\D^{l}_{\beta\rho}(y,x)\,\D^{l+k}_{\sigma\alpha}(x,y)\,\left[\eta^{\nu\alpha} (K+L')^{\beta}-\eta^{\alpha\beta} (L'+L)^{\nu} +\eta^{\beta\nu})(L-K)^{\alpha}\right](y)\,.
\mathcal{M} = \left[\eta^{\mu\rho} (K-L)^{\sigma}+\eta^{\rho\sigma} (L+L')^{\mu}-\eta^{\sigma\mu} (L'+K)^{\rho}\right]\!(x)\, \cE_{\mu}(x)\,\cE'_{\nu}(y) \\
\D^{l}_{\beta\rho}(y,x)\,\D^{l+k}_{\sigma\alpha}(x,y)\,
\left[-\eta^{\nu\alpha} (K+L')^{\beta}+\eta^{\alpha\beta} (L'+L)^{\nu} -\eta^{\beta\nu}(L-K)^{\alpha}\right]\!(y)\,.
\end{multline}
In this expression, the action of the covariant derivatives in the cubic gluon vertices \eqref{gl3pv} leads directly to dressed momenta insertions: $D_{j\,\mu}\rightarrow \im K_{j\,\mu}$ for leg $j$ of the vertex. While this is what one might expect from the vertex on a flat background, it is not \emph{a priori} obvious that this should be the case, as the covariant derivatives on the plane wave background can also act on the tensor structures outside of the exponential. Nevertheless, one finds that the derivative contributions from the tensor structure only produce terms (proportional to $n_\mu$) which are ultimately contracted to zero with other contributions to $\mathcal{M}$.
   
The expression \eqref{swamp1} is to be evaluated on the support which follows from the loop integrations, so that $L'$ is replaced by \eqref{l-replace-2}.  The terms proportional to $n_\mu$ in \eqref{l-replace-2} ultimately vanish, which yields some simplifications, as does dropping terms proportional to $\cE\cdot\cE'$ which will not contribute to helicity flip. 

To present the result we introduce the following notation. Contractions of two generic vectors $U^\mu$ and $V^\nu$ via the gluon propagator carrying momentum $p_\mu$ will be denoted
\be\label{notation1}
	(U|p|V) := U^{\mu}(x)\,\D^{p}_{\mu\nu}(x,y)\,V^{\nu}(y)  \;,
%K^{\mu}(x)\,\D^{p}_{\mu\nu}(x,y)\,L^{\nu}(y):=(K|p|L)\,,
% \qquad \cE^{\mu}(x)\,\D^{p}_{\mu\nu}(x,y)\,L^{\nu}(y):=(\cE|p|L)\,,
\ee
while contractions via two gluon propagators will be written
\be\label{notation2}
	%\cE_{\mu}[p|q]:=\cE^{\sigma}(x)\,\D^{p\,\,\alpha}_{\sigma}(x,y)\,\D^{q}_{\alpha\mu}(y,x)\,.
	(U|p,q|V)_x := U^\sigma(x)\,\D^{p\,\,\alpha}_{\sigma}(x,y)\,\D^{q}_{\alpha\mu}(y,x) V^\mu(x) \;,
\ee
in which an extra position subscript is needed. Using this notation, the gluon loop contributes
\begin{multline}\label{gll1}
%(\cE|l|2K+L)\,(K-L|l+k|\cE') + (\cE|l+k|K-L)\,(2K+L|l|\cE')-4d\,\cE\cdot L(x)\,\cE'\cdot L(y) \\
%+2\cE'\cdot L(y)\,\Big[ (\cE|l+k,l | 2K+L)_x\,-(\cE|l,l+k|K-L)_x \Big] \\
%+2 \cE\cdot L(x)\,\Big[ (\cE'|l+k,l | 2K+L)_y\,-(\cE'|l,l+k|K-L)_y \Big] \,. %\Anton{\text{All agreed!}}
\mathcal{M} = 4d\,\cE\cdot L(x)\,\cE'\cdot L(y) -(\cE|l|2K+L)\,(K-L|l+k|\cE') - (\cE|l+k|K-L)\,(2K+L|l|\cE') \\
-2\cE'\cdot L(y)\,\Big[ (\cE|l+k,l | 2K+L)_x\,-(\cE|l,l+k|K-L)_x \Big] \\
-2 \cE\cdot L(x)\,\Big[ (\cE'|l+k,l | 2K+L)_y\,-(\cE'|l,l+k|K-L)_y \Big] \,. %\Anton{\text{All agreed!}}
\end{multline}
This is symmetric between $x\leftrightarrow y$ and $\cE\leftrightarrow\cE'$ and in the flat background limit is easily seen to reduce to $(4d-6) \varepsilon\cdot l \varepsilon'\cdot l$, as expected for this diagram.

\subsubsection{Gluon tadpole}
In the tadpole diagram all propagators and external legs meet at the same point; this results in all dependence on the plane wave background dropping out of the calculation. The whole diagram is then exactly equal to the flat-background result. In this case the lightfront time integrals can be performed, giving another momentum-conserving delta function, and so (neglecting numerical factors) the gluon tadpole contributes
\be
	M_\text{tadpole}(k) \sim (d-1)g^2 C_2(G)\, \delta^{\mathsf{ab}}\,\delta(k_\LCm+k'_\LCm)\, \varepsilon\cdot \varepsilon^\prime\, \oint\!\frac{\ud^d l}{l^2+i\epsilon} \,.
\ee
The contraction of the polarization vectors means that the whole expression vanishes for helicity flip, while for non-flip it would be subtracted entirely as part of the renormalisation. The tadpole contribution can therefore be neglected.

\subsubsection{Quark loop}

In the fermionic quark loop, there is a gamma matrix trace to be calculated which takes the form
\be\label{the-trace}
	\tr \slashed{\cE}(x)\, V^l(x,y)\, \slashed{\cE}'(y)\, V^{l'}(y,x) \;. %\to \frac{1}{4n.l n.l'}\tr \slashed{\cE}(x) \slashed{L}(x)\slashed{n}\slashed{L}(y) \slashed{\cE}(y) \slashed{L'}(y)\slashed{n}\slashed{L'}(x)  
\ee
Recall that this trace is to be evaluated on the support of the delta functions resulting from the position and momentum integrals above, and that we have identified in \eqref{VolkNew} the representation $V^l = \vec{V}^L + I^l$. Now, terms coming from $I^{l}$ behave differently from the others, and from those considered above, as such terms contribute to the propagator
\be
	\frac{1}{l^2-m^2}\, I^l  = \frac{\slashed{n}}{2\,l_\LCp} \;.
\ee%
This is not only independent of the coordinates, but also removes the pole. It can easily be seen from \eqref{the-trace}, and using $\varepsilon\cdot\varepsilon' = 0 = n\cdot\varepsilon$, that such instantaneous propagator terms do not contribute to helicity flip at one loop. (The same is true in QED using both Feynman diagram and lightfront Hamiltonian methods~\cite{Dinu:2013gaa}). 

Turning to $\vec{V}^{L}(x,y)$, observe that this part of the propagator is essentially `on-shell,' because \eqref{VolkNew} is the background-dependent generalization of the vacuum identity
\be\begin{split}
	\frac{-\slashed{l}+m}{l^2-m^2}  &= \frac{-\slashed{l}^\text{o.s.}+m}{l^2-m^2} + \frac{\slashed{n}}{2\,l_\LCp} =   -\frac{1}{2\,l_\LCp}\, \frac{(\slashed{l}-m) \slashed{n} (\slashed{l} -m )}{l^2-m^2} + \frac{\slashed{n}}{2\,l_\LCp} \;,
\end{split}
\ee
where $l^\text{o.s.}_\mu$ is the momentum put on-shell by fixing the $n_\mu$ component. In particular, this means that $\vec{V}^{L'}$ can be replaced by $\vec{V}^{L+K}$ thanks to the factor of $\slashed{n}$.

The trace \eqref{the-trace} therefore reduces to
\be\label{TrTr}
	\tr \big[ \slashed{\cE}(x)\, {\vec V}^L(x,y)\, \slashed{\cE}'(y)\, {\vec V}^{L+K}(y,x) \big] %\to \tr \big[ \slashed{\cE}(x) \slashed{L}(x)\slashed{n}\slashed{L}(y)\slashed{\cE}(y) {\vec V}_{L+K}(y,x) \big]
\ee
%
%where, in the second simplification and as can be verified, the explicit mass terms do not contribute to helicity flip.
%
From this point, the trace is calculated as normal. It is helpful in the calculation to note the following results. First, the \emph{explicit} mass terms in $\vec V$ do not contribute. Second, we are interested in helicity flip, for which $\varepsilon\cdot\varepsilon'=0$; it is easy to show that this orthogonality holds even when dressed polarizations are contracted at \textit{different} lightfront times:
\be
	\cE(x^\LCm)\cdot\cE'(y^\LCm) = 0 \;. 
\ee
Third, 
\be\label{transport}
	\cE(x^\LCm)\cdot\big(L(y^\LCm) + s\, K(y^\LCm)\big) = \cE\cdot L(y^\LCm) \;,
\ee
in which, as above, $s\equiv -l_{\LCp}/k_\LCp$. This shows that the combination $L+sK$ -- which appears when calculating the trace -- acts as a kind of transport, pulling the polarization vector it is contracted with back to its own lightfont time.

The final result is, for each flavour of quark, %for $n_f$ the number of flavours,
\be\begin{split}\label{quark-final-1}
	M_\text{quark}(k) = %-2 \frac{g^2 T_F n_f
	-2 \frac{g^2 T_F \delta^{\mathsf{ab}}}{k_\LCp (2\pi)^{d-1}} &\int\limits_{-\infty}^{+\infty}\!\ud x^\LCm\!\int\limits^{x^\LCm}_{-\infty}\ud y^\LCm\! 
	\int\limits_0^1\!\frac{\ud s}{s(1-s)} \int\!\ud^{d-2}l_\LCperp
	\, \e^{ \frac{\im}{k_{\LCp}(1-s)}\int\limits_{y^\LCm}^{x^\LCm}\!\ud\alpha\, K\cdot \tilde{L}(\alpha)} \\
		&\times \left[ \mathscr{S}\cE'\cdot\tilde{L}(y) \cE\cdot\tilde{L}(x) + \bigg(1-\frac{1}{2s(1-s)}\bigg) \mathscr{A}\cE'\cdot\tilde{L}(y) \cE\cdot\tilde{L}(x)  \right] \;,
%	&\times (-1)\, 8\cE^{\prime \mu}(y)\cE^\nu(x)\bigg[ J_{(\mu,}(x) J_{\nu)}(y) + \bigg(1-\frac{1}{2s(1-s)}\bigg) J_{[\mu,}(x) J_{\nu]}(y)  \bigg] \;,
\end{split}
\ee
where $T_F$ is defined by $\mathrm{tr}(\mathsf{T}^{\sa}\mathsf{T}^{\mathsf{b}})=T_F \delta^{\mathsf{ab}}$ and there is an overall factor of $-8$ relative to \eqref{gluon-final-1} coming from the fermion loop and the trace \eqref{TrTr}. The symbol ($\mathscr{A}$) $\mathscr{S}$ means (anti) symmetrise in $y^\LCm$ and $x^\LCm$ (with factors of 1/2). The loop momenta $\tilde{L}$ are decorated with a tilde to remind us that they are \emph{massive} momenta, rather than the massless momenta appearing in the gluon and ghost loops. It is easy to see that in the flat background limit the trace reduces to
\be
\tr \big[\slashed{\varepsilon}\slashed{l} \slashed{\varepsilon'} \slashed{l}\big] = 8 \,\varepsilon\cdot l\, \varepsilon'\cdot l \;,
\ee
which is again the correct limit for this diagram.  We remark that the trace is obtained from the QED result by applying the two rules: 1.) replace the free photon polarization vectors with the dressed gluon polarization vectors, and 2.) evaluate the dressed polarizations at the same point as the dressed momenta they are contracted with. That this second rule should hold is not obvious from the QED result, where the photon polarization vectors are independent of position, but is affected by the transport property (\ref{transport}).

%%%%%%%%%%%%%%%%%%%%%%%%%%%%%%%%%%%%%%%%

\subsection{Helicity flip amplitude}

At this stage, we have all of the ingredients to assemble the full helicity flip amplitudes for pure Yang-Mills and for QCD, with $N$ colours and $n_f$ flavours. In general, amplitudes contain divergences in $d=4$; these may be regulated using transverse dimensional regularisation~\cite{Casher:1976ae} with respect to $\d^{d-2}l_{\LCperp}$. As this corresponds to continuing the number of transverse space-time directions, it is trivial to similarly continue the background away from $d=4$. Here, one can show that all terms which depend on the background field are UV finite (cf., \cite{Becker:1974en,Dinu:2013gaa} for analogous statements in QED), so the renormalized amplitude is given by subtracting off the flat background contribution~\cite{Dittrich:2000zu}. 

However, it is easy to see that the flat background amplitude is proportional to $\varepsilon\cdot\varepsilon'$, which vanishes in the case of helicity flip. Therefore, the relevant helicity flip amplitude for pure Yang-Mills is:
\be\label{YMdhf}
\boxed{M^{\mathrm{YM}}(k)=M_{\text{ghost}}(k)+M_{\text{gluon}}(k)\,.}
\ee
Note, as observed above, that $M_{\text{tadpole}}$ does not contribute to helicity flip, being equal to its flat-background value. Similarly, the QCD helicity flip amplitude is given by:
\be\label{QCDdhf}
\boxed{M^{\mathrm{QCD}}(k)=M^{\mathrm{YM}}(k)+\sum_{i=1}^{n_f}M^{(i)}_{\text{quark}}(k)\,,}
\ee
where $M^{(i)}_{\text{quark}}$ is the quark loop contribution for the $i^{\mathrm{th}}$ flavour.

In order to obtain more explicit formulae for the helicity flip amplitude we now take explicit choices for the polarization of the incoming gluon. As we will see, particularly clean expressions can be obtained by making use of special representations available for on-shell kinematics in $d=4$, even in the presence of a background plane wave.

%\Anton{
%Using Peskin, and Schwartz, I seem to find for the relative contributions
%
%\be
%	C_2(G) (2d-4) - 8 T_F n_f
%\ee
%
%}

%%%%%%%%%%%%%%%%%%%%%%%%%%%%%%%%%%%%%%%%%%%%%%%%
%%%%%%%%%%%%%%%%%%%%%%%%%%%%%%%%%%%%%%%%%%%%%%%%

\section{Spinor-helicity formalism}
\label{SHF}

One of the key tools in the modern approach to scattering amplitudes is the \emph{spinor helicity formalism}, which enables streamlined representations of on-shell kinematic data in $d=4$ space-time dimensions (cf., \cite{Srednicki:2007qs,Elvang:2013cua,Dixon:2013uaa,Cheung:2017pzi}; our conventions follow~\cite{Adamo:2017qyl}). At the heart of this formalism is the isomorphism between the complexified Lorentz group SO$(4,\C)$ and SL$(2,\C)\times\mathrm{SL}(2,\C)$, as is realized by the Pauli matrices $\sigma_{\mu}^{\alpha\dot\alpha}$: contraction with the Pauli matrices enables any Lorentz index to be interchanged for a pair of SL$(2,\C)$ Weyl spinor indices: $v^{\mu}\leftrightarrow v^{\alpha\dot\alpha}=v^{\mu}\sigma_{\mu}^{\alpha\dot\alpha}$. 

This spinor notation is particularly useful when considering an on-shell 4-vector $k^{\mu}$, obeying $k^2=0$. In the spinor notation, it is easy to see that $k^2=0$ is equivalent to $\mathrm{det}(k^{\alpha\dot\alpha})=0$, which implies that $k^{\alpha\dot\alpha}$ must be a simple $2\times 2$ matrix: $k^{\alpha\dot\alpha}=\lambda^{\alpha}\bar{\lambda}^{\dot\alpha}$. Conversely, any 4-vector which is representable as the product of two spinors must be null. Therefore it follows that
\be\label{fsh1}
k^2=0\:\Leftrightarrow \:k^{\alpha\dot\alpha}=\lambda^{\alpha}\,\bar{\lambda}^{\dot\alpha}\,,
\ee
where reality of $k^{\mu}$ dictates that $\lambda^{\alpha}$, $\bar{\lambda}^{\dot\alpha}$ are related by complex conjugation.

One could worry that the presence of background fields spoils this statement (and others which follow from it) in some way, but the spinor helicity formalism extends naturally to on-shell fields on the plane wave background. We demonstrate how to represent the dressed gluon momenta and pure positive/negative helicity polarization vectors in this dressed version of the spinor helicity formalism. To our knowledge, this is the first time that the formalism has been adapted to plane wave background fields, and we expect that it should be a useful tool beyond the helicity flip calculations which are the focus of this paper.

\medskip

It will be useful to write the $d=4$ Minkowski metric in lightfront coordinates which represent the transverse $x^\LCperp$-directions as the complex plane:
\be\label{Minkmet}
\d s^2 = 2\left(\d x^{+}\,\d x^{-}-\d z\,\d\bar{z}\right)\,.
\ee
Space-time coordinates in the spinor helicity formalism are now encoded in the $2\times2$ matrix:
\be\label{scoords}
x^{\alpha\dot\alpha}=\frac{1}{\sqrt{2}}\left(\begin{array}{cc}
                                                                     x^{0}+x^{3} & x^{1}-\im x^{2} \\
                                                                     x^{1}+\im x^2 & x^0-x^3
                                                                     \end{array}\right) = \left(\begin{array}{cc}
                                                                                                             x^+ & \bar{z} \\
                                                                                                             z & x^- 
                                                                                                             \end{array}\right)\,,
\ee
with $x^{\LCperp}=(x^1,x^2)$ replaced by $(z,\bar{z})$. From now on, all expressions will be given in these coordinates. The collection of non-trivial background field degrees of freedom, $a_{\LCperp}(x^\LCm)$, are repackaged as:
\be\label{shbf}
a(x^\LCm):=\frac{a_1(x^\LCm)+\im\,a_2(x^\LCm)}{\sqrt{2}}\,, \qquad \bar{a}(x^\LCm):=\frac{a_1(x^\LCm)-\im\,a_2(x^\LCm)}{\sqrt{2}}\,.
\ee
Spinor indices are raised and lowered with the Levi-Civita symbols 
\be\label{Levi-Civita}
\epsilon_{\alpha\beta}=\left(\begin{array}{c c}
                              0 & 1 \\
                              -1 & 0
                             \end{array}\right) = \epsilon_{\dot{\alpha}\dot{\beta}}\,,
\ee
and their inverses according to the convention
\begin{equation*}
a_{\alpha}:=a^{\beta}\,\epsilon_{\beta\alpha}\,, \qquad b^{\alpha}:=\epsilon^{\alpha\beta}\,b_{\beta}\,,
\end{equation*}
and so forth.

The lightlike vector $n^{\mu}$ associated with the background plane wave has the spinor expression
\be\label{lcv1}
n^{\alpha\dot\alpha}=\left(\begin{array}{cc}
                                          1 & 0 \\
                                          0 & 0
                                          \end{array}\right)\,.
\ee
Since $n^2=0$, we can write $n^{\alpha\dot\alpha}$ as a product of spinors:
\be\label{lcv2}
n^{\alpha\dot\alpha}=n^{\alpha}\,\bar{n}^{\dot\alpha}\,, \qquad n^{\alpha}=\left(\begin{array}{c}
1 \\
0
\end{array}\right)\,.
\ee
Similarly, the on-shell dressed gluon momentum $K_\mu(x^{\LCm})$ of \eqref{K-DEF} obeys $K^2=0$ and should therefore also be expressible as a product of spinors. In the 4d lightfront coordinates \eqref{Minkmet}, the dressed momentum $K_{\alpha\dot\alpha}$ is:
\be\label{smom1}
K_{\alpha\dot\alpha}(x^\LCm)=\left(\begin{array}{cc}
                                          k_\LCp & \bar{k}+e\, \bar{a} \\
                                          k+e\, a & \frac{|k+e\, a|^2}{k_\LCp}
                                          \end{array}\right)\,,
\ee
where $k=(k_{1}+\im k_{2})/\sqrt{2}$ and $\bar{k}=(k_{1}-\im k_{2})/\sqrt{2}$ are the transverse un-dressed momenta in complex coordinates. The decomposition of $K_{\alpha\dot\alpha}$ into spinors is given by
\be\label{smom2}
K_{\alpha\dot\alpha}=\Lambda_{\alpha}\,\bar{\Lambda}_{\dot\alpha}\,, \qquad \Lambda_{\alpha}=\left(\begin{array}{c} \sqrt{k_\LCp} \\ \frac{k+e\, a}{\sqrt{k_\LCp}} \end{array}\right)\,, \quad
\bar{\Lambda}_{\dot\alpha}=\left(\begin{array}{c} \sqrt{k_\LCp} \\ \frac{\bar{k}+e\,\bar{a}}{\sqrt{k_\LCp}} \end{array}\right)\,.
\ee
The only difference between \eqref{smom2} and the decomposition $k_{\alpha\dot\alpha}=\lambda_{\alpha}\bar{\lambda}_{\dot\alpha}$ on a flat background is that the spinors $\Lambda_{\alpha}$, $\bar{\Lambda}_{\dot\alpha}$ are non-trivial functions of the lightfront coordinate $x^\LCm$ through dependence on the background field components $a(x^\LCm)$, $\bar{a}(x^\LCm)$. This is nothing but a boost applied to the free spinors, as follows from \eqref{K-DEF2}.  

\medskip

A key advantage of the spinor helicity formalism is that it allows gluon polarizations to be replaced with a positive or negative helicity label. Indeed, given an on-shell momentum and this helicity label, the corresponding polarization vector is uniquely specified up to gauge redundancy (cf., \cite{Witten:2003nn}). The same is true for gluons in the plane wave background, and their dressed polarization vectors. 

A general dressed polarization vector is written in spinor components as
\be\label{spol1}
\cE_{\alpha\dot\alpha}(x^\LCm)=\left(\begin{array}{cc}
                                             0 & \epsilon \\
                                             \bar{\epsilon} & \frac{\epsilon (\bar{k}+e\,\bar{a})+\bar{\epsilon} (k+e\,a)}{k_\LCp}
                                             \end{array}\right)\,,
\ee                                             
with the two on-shell degrees of freedom parametrized by $\{\epsilon,\bar{\epsilon}\}$. From this, the purely negative and positive helicity polarization states are:
\be\label{spol2}
\cE^{(-)}_{\alpha\dot\alpha}=\frac{\Lambda_{\alpha}\,\bar{n}_{\dot\alpha}}{\sqrt{k_\LCp}}\,, \qquad \cE^{(+)}_{\alpha\dot\alpha}=\frac{n_{\alpha}\,\bar{\Lambda}_{\dot\alpha}}{\sqrt{k_\LCp}}\,.
\ee
Note that these polarization vectors are clearly transverse, $\cE^{(-)}\cdot K=0=\cE^{(+)}\cdot K$, and compatible with lightcone gauge\footnote{The residual gauge freedom familiar from the flat background spinor helicity formalism is fixed in \eqref{spol2} by the additional requirement of lightcone gauge.}, $\cE^{(-)}\cdot n=0=\cE^{(+)}\cdot n$.

It is straightforward to confirm that the assignments of negative/positive helicity to the polarizations of \eqref{spol2} is correct. Indeed, defining the gluon wavefunction $a_{\alpha\dot\alpha}^{(-)}=\cE^{(-)}_{\alpha\dot\alpha} \mathrm{e}^{\im\,\phi_k}$, its linearised field strength is:
\be\label{nhfs}
f^{(-)}_{\alpha\dot\alpha\beta\dot\beta}=2\,D_{[\alpha\dot\alpha} a_{\beta\dot\beta]}^{(-)}=\frac{\im}{2}\,\epsilon_{\dot\alpha\dot\beta}\,\Lambda_{\alpha}\Lambda_{\beta}\,\mathrm{e}^{\im\,\phi_k}\,,
\ee
as expected for a negative helicity state. Similarly,
\be\label{phfs}
f^{(+)}_{\alpha\dot\alpha\beta\dot\beta}=\frac{\im}{2}\,\epsilon_{\alpha\beta}\,\bar{\Lambda}_{\dot\alpha}\bar{\Lambda}_{\dot\beta}\,\mathrm{e}^{\im\,\phi_k}\,,
\ee
for the positive helicity polarization.

The power of these dressed helicity polarizations is that they enable us to consider the `pure' helicity flip amplitude, which substantially simplifies the integrands appearing in \eqref{YMdhf} and \eqref{QCDdhf}.

%%%%%%%%%%%%%%%%%%%%%%%%%%%%%%%%%%%%%%%%%%%%%%%%%%%%%
%%%%%%%%%%%%%%%%%%%%%%%%%%%%%%%%%%%%%%%%%%%%%%%%%%%%%

\section{Negative to positive helicity flip}
\label{GHF2}

Armed with the spinor helicity formalism, we now consider the specific case of negative-to-positive helicity flip. The negative helicity incoming gluon has the polarization vector $\cE^{(-)}_{\alpha\dot\alpha}$ in \eqref{spol2}. Because the polarization vector of the outgoing, positive helicity gluon obeys $\overline{\cE'}_{\alpha\dot\alpha}=\cE^{(+)}_{\alpha\dot\alpha}$, we set $\cE'_{\alpha\dot\alpha}=\cE^{(-)}_{\alpha\dot\alpha}$ in our amplitudes $M(k)$.

The amplitudes \eqref{YMdhf} and \eqref{QCDdhf} are now evaluated using the explicit expressions for $\cE^{(-)}_{\alpha\dot\alpha}$ and $K_{\alpha\dot\alpha}$ in the spinor helicity formalism. It is also useful to have the spinor parametrizations of the loop momenta; for all diagrams contributing to the pure Yang-Mills result this is:
\be\label{loopints}
L_{\alpha\dot\alpha}=\left(\begin{array}{cc}
                                    l_{\LCp} & \bar{l}+e_l\, \bar{a} \\
                                    l+e_l\, a & \frac{|l+e\, a|^2}{l_\LCp}
                                    \end{array}\right)\,,                                   
\ee
while for the quark loop the momentum is
\be\label{qloopints}
\tilde{L}_{\alpha\dot\alpha}=\left(\begin{array}{cc}
                                    l_{\LCp} & \bar{l}+\mu_l\, \bar{a} \\
                                    l+\mu_l\, a & \frac{|l+\mu_l\, a|^2+m^2}{l_\LCp}
                                    \end{array}\right)\,.
\ee                                    
Computation of the integrands appearing in \eqref{YMdhf} and \eqref{QCDdhf} is now a simple matter of spinor helicity manipulations and (mostly) Gaussian integration.

%%%%%%%%%%%%%%%%%%%%%%%%%%%%%%%%%%%%%%%%%%%%%%%%%%%%%

\subsection{Pure Yang-Mills}

A straightforward calculation using the spinor helicity formalism shows that the non-trivial integrand structures appearing in the ghost and gluon loop diagrams are in fact proportional:
\be\begin{split}
	\text{ghost} \quad &(\ref{ghost-final-1}) \longrightarrow -\cE^{(-)}\cdot L(x^{\LCm})\,\cE^{(-)}\cdot L(y^{\LCm})\,,  \\
	\text{gluon} \quad &(\ref{gluon-final-1}) \longrightarrow (2d-3)\,\cE^{(-)}\cdot L(x^\LCm)\,\cE^{(-)}\cdot L(y^\LCm)\,,
\end{split}
\ee
with the quantity
\be\label{int1}
\cE^{(-)}\cdot L(x^{\LCm})\,\cE^{(-)}\cdot L(y^{\LCm})=\left(l+s\,k+(e_l+s\,e)\,a(x^\LCm)\right)\,\left(l+s\,k+(e_l+s\,e)\,a(y^\LCm)\right)\,.
\ee
Note that only the `holomorphic' components of the background gauge field appear in this expression; this is due to the pure negative helicity of the incoming gluon.

The pure Yang-Mills amplitude then takes the form
\be\begin{split}\label{YMhf1}
\frac{g^2\,N\,\delta^{\mathsf{ab}}}{k_\LCp\, (2\pi)^{3}} &\int\limits_{-\infty}^{+\infty}\!\ud x^\LCm\!\int\limits^{x^\LCm}_{-\infty}\ud y^\LCm\! 
	\int\limits_0^1\!\frac{\ud s}{s(1-s)} \int\!\ud^{2}l
	\, \e^{ \frac{\im}{k_{\LCp}(1-s)}\int\limits_{y^\LCm}^{x^\LCm}\!\ud\alpha\, K\cdot L(\alpha)} \\
%	&\times\Bigg[ \left(l+s\,k+(e_l+s\,e)\,a(x^\LCm)\right)\,\left(l+s\,k+(e_l+s\,e)\,a(y^\LCm)\right)-(l+s\,k)^2\Bigg]\,,
	&\times\Big[ \left(l+s\,k+(e_l+s\,e)\,a(x^\LCm)\right)\,\left(l+s\,k+(e_l+s\,e)\,a(y^\LCm)\right)\Big]\,,
\end{split}
\ee
where we have used $C_2(G)=N$ for $G=\mathrm{SU}(N)$.

To deal with the exponential in the integrand of \eqref{YMhf1}, we write $\phi:=(x^\LCm + y^\LCm)/2$, $\theta:=x^{\LCm}-y^{\LCm}$, and introduce the moving average
\be\label{maverage}
	\la a\ra:=\frac{1}{\theta}\int\limits_{\phi-\theta/2}^{\phi+\theta/2}\!\d\alpha\, a(\alpha)\,.
\ee
With this, we note that the exponent is extremised at
\be\label{vexp1}
l^{\star}=-e_l\,\la a\ra-s\,\left(k+e\,\la a\ra\right)\,, \qquad \bar{l}^{\star}=-e_l\,\la \bar{a}\ra-s\,\left(\bar{k}+e\,\la \bar{a}\ra\right)\,.
\ee
Changing variables in the transverse loop momenta from $l$ to $q= l- l^{\star}$ allows us to write the exponent as
\be\label{vexp2}
\frac{-\im\,\theta}{k_{+}\,s\,(1-s)}\, \big(q\bar{q}+\mathrm{var}(a)\,(e_l+s\,e)^2\big)\,,
\ee
where the floating variance is defined by~\cite{Kibble:1965zza,Kibble:1975vz,Hebenstreit:2010cc,Harvey:2012ie}
\be\label{fvariance}
	\mathrm{var}(a):=\left\la a \bar{a}\right\ra-\la a\ra\, \la {\bar a}\ra\,.
\ee
Note that setting $e=0$ in \eqref{vexp2} recovers the standard Volkov exponent from QED.

The Gaussian integration in $\{q,\bar{q}\}$ can now be performed, resulting in %and subtracting the flat background counterterm results in:
\be\begin{split}\label{MYM}
M^{\text{YM}}_{-\rightarrow +}(k) &= \frac{-\im\,g^2\,N\, \delta^{\mathsf{ab}}}{(2\pi)^{2}} \int\limits_{-\infty}^{+\infty}\!\ud\phi \!\int\limits_0^\infty\! \frac{\ud\theta}{\theta}\!\int\limits_0^1\!\ud s 
  \,\exp\!\bigg[- \im \frac{\theta\,\text{var}(a)}{s(1-s)k_\LCp} \,(e_l +se)^2\bigg] \\
	& \times(e_l+ se)^2 \big({a}(\phi+\theta/2)- \langle {a} \rangle\big)\, \big({a}(\phi-\theta/2)- \langle {a} \rangle\big) \,.
\end{split}
\ee
The factor of $\theta^{-1}$ appearing in \eqref{MYM} does not lead to a UV divergence, because the remainder of the integrand goes like $\sim\theta^2$ at small $\theta$, such that the integrand vanishes linearly in the small $\theta$ region. Note that for general gluon polarizations, there would be $\theta^{-2}$ factors appearing in the integrand; the softer UV behaviour of $\theta^{-1}$ in \eqref{MYM} is due to the choice of a pure negative helicity polarization for the incoming gluon. As a result, the non-exponential integrand in \eqref{YMhf1} is a (holomorphic) function of $l$ (or $q$); for a general linear combination of helicity states the integrand can be a quadratic polynomial in both $l$ and $\bar{l}$, leading to factors of $\theta^{-2}$ after the Gaussian integration is performed. 

The fact that the helicity flip integral is well-behaved is easily seen by defining the well-behaved functions $a_\theta:= \partial_\theta \langle a\rangle$ and $a_\phi:= \partial_\phi \langle a\rangle$, from which it can be seen that
\be\label{urgh}
	 \big(a(\phi+\theta/2)- \langle a \rangle\big) \big(a(\phi-\theta/2)- \langle a \rangle\big) = \theta^2\,\big( a_\theta a_\theta  -\frac{1}{4}a_\phi  a_\phi \big)\,.
\ee
Even with general gluon polarizations, it is easy to see that the amplitude is convergent. Finally, the helicity flip amplitude for pure Yang-Mills can be written as
\be\label{MYM2}\boxed{\begin{aligned}
M^{\text{YM}}_{-\rightarrow +}(k) &= \frac{-\im\,g^2\,N}{(2\pi)^{2}} \sum_{e_l} \int\limits_{-\infty}^{+\infty}\!\ud\phi \!\int\limits_0^\infty\!\ud\theta\,\theta\!\int\limits_0^1\!\ud s 
  \,\exp\!\bigg[- \im \frac{\theta\,\text{var}(a)}{s(1-s)k_\LCp} \,(e_l +s e)^2\bigg] \\
	& \times(e_l+ se)^2  \left( a_\theta a_\theta  -\frac{1}{4}a_\phi  a_\phi \right)\,,
   \end{aligned}}
\ee
where we suppress the trivial colour structure $\delta^{\mathsf{ab}}$, and the sum is over all (adjoint) charges $e_l$ flowing in the loop which are consistent with charge conservation.

%%%%%%%%%%%%%%%%%%%%%%%%%%%%%%%%%%%%%%%%%%%%%%%%%%%%%

\subsection{QCD}

In the case of the pure negative to positive helicity flip, only the symmetric structure from \eqref{quark-final-1} contributes to the amplitude. Thus, the only contribution to the integrand in the quark diagram reads: 
\be\begin{split}
\text{quarks} \quad &(\ref{quark-final-1}) \longrightarrow -8\, \cE^{(-)}\cdot \tilde{L}(x^\LCm)\, \cE^{(-)}\cdot \tilde{L}(y^\LCm)\,.
\end{split}
\ee
Each flavour of quark then contributes %For QCD with $n_f$ flavours, the quark loops thus contribute
\be\begin{split}\label{QCDhf1}
%-\frac{g^2\,n_f\,\delta^{\mathsf{ab}}}{k_\LCp\, (2\pi)^{3}}
-\frac{g^2\,\delta^{\mathsf{ab}}}{k_\LCp\, (2\pi)^{3}} &\int\limits_{-\infty}^{+\infty}\!\ud x^\LCm\!\int\limits^{x^\LCm}_{-\infty}\ud y^\LCm\! 
	\int\limits_0^1\!\frac{\ud s}{s(1-s)} \int\!\ud^{2}l
	\, \e^{ \frac{\im}{k_{\LCp}(1-s)}\int\limits_{y^\LCm}^{x^\LCm}\!\ud\alpha\, K\cdot \tilde{L}(\alpha)} \\
%	&\times\Bigg[ \left(l+s\,k+(g\mu_l+s\,e)\,a(x^\LCm)\right)\,\left(l+s\,k+(g\mu_l+s\,e)\,a(y^\LCm)\right)-(l+s\,k)^2\Bigg]\,,
	&\times\Big[ \left(l+s\,k+(\mu_l+s\,e)\,a(x^\LCm)\right)\,\left(l+s\,k+(\mu_l+s\,e)\,a(y^\LCm)\right)\Big]\,,
\end{split}
\ee
%with the UV divergent flat background contribution subtracted in the final line.
At this point, the Gaussian integration over the transverse loop momenta can be performed as in pure Yang-Mills; the only difference is that the argument of the exponential takes the form:
\be\label{quarkexp}
	\frac{-\im\,\theta}{k_{+}\,s\,(1-s)}\, \big(q\bar{q}+\frac{m_i^2}{2}+(\mu_l+s\,e)^2\mathrm{var}(a)\big)\,,
\ee
where $m_i$ is the mass of the $i^{\mathrm{th}}$ quark flavour. When $e=0$ (and setting $\mu_l\to 1$) this is precisely the standard Volkov exponent appearing in QED. It is useful to introduce the notation
\be\label{effmass}
	\mathrm{M}^2_i[\mu_l]:=m_i^2+2(\mu_l+s\, e)^2\mathrm{var}(a)\,,
\ee
for the effective quark mass, which trivially generalizes Kibble's effective mass for the electron in strong field QED~\cite{Kibble:1975vz}.

Performing the Gaussian integrations gives the QCD negative-to-positive helicity flip amplitude
\be\label{QCDhf2}\boxed{\begin{aligned}
 M^{\text{QCD}}_{-\rightarrow +}(k) &= \frac{\im\,g^2}{(2\pi)^2} \sum_{i=1}^{n_f} \sum_{\mu_l} \int\limits_{-\infty}^{+\infty}\!\ud\phi \!\int\limits_0^\infty\!\ud\theta\,\theta\!\int\limits_0^1\!\ud s 
  \,\exp\!\bigg[- \im \frac{\theta\, \mathrm{M}_i^2[\mu_l]}{2s(1-s)k_\LCp} \bigg] \\
	& \times(\mu_l+ se)^2  \left( a_\theta a_\theta  -\frac{1}{4}a_\phi  a_\phi \right) \:\: +\, M^{\text{YM}}_{-\rightarrow +}(k)\,,                     
                        \end{aligned}}
\ee                        
where the sums are over all flavours $i$ and over all fundamental weights compatible with charge conservation which flow in the loop. Once again, the integrand is regular in $\theta$; there is no UV divergence.

%%%%%%%%%%%%%%%%%%%%%%%%%%%%%%%%%%%%%%%%%
%%%%%%%%%%%%%%%%%%%%%%%%%%%%%%%%%%%%%%%%%

\section{Examples}
\label{Exam}

The formulae \eqref{MYM2}, \eqref{QCDhf2} still contain three residual integrals: two over the lightfront directions $(\phi,\theta)$ and the $s$ integral corresponding to the $l_{\LCp}$ loop momentum component. The lightfront integrals are a general feature of calculations in a plane wave background, while the $s$ integral can only be performed if the charges flowing in the loop are explicitly known. In this section, we consider some concrete examples and limits of the helicity flip amplitude which allow some or all of these residual integrals to be performed analytically.

%%%%%%%%%%%%%%%%%%%%%%%%%%%%%%%%%%%%%%%%%

\subsection{Gauge group SU(2)}

Consider a gauge theory with $N=2$ colours; the generators of the gauge group SU$(2)$ can be written in terms of the familiar basis
\be\label{su21}
\mathsf{T}^{+}=\left(\begin{array}{cc}
                      0 & 1 \\
                      0 & 0
                     \end{array}\right)\,, \qquad \mathsf{T}^{-}=\left(\begin{array}{cc}
                                                                        0 & 0 \\
                                                                        1 & 0
                                                                       \end{array}\right)\,, \qquad \mathsf{T}^{0}=\left(\begin{array}{cc}
                                                                                                                          \frac{1}{2} & 0 \\
                                                                                                                          0 & -\frac{1}{2}
                                                                                                                         \end{array}\right)\,,
\ee
where $\mathsf{T}^0$ is the generator of the Cartan. Since 
\be\label{adjcharge}
\left[\mathsf{T}^{0},\,\mathsf{T}^{\pm}\right]=\pm \mathsf{T}^{\pm}\,,
\ee
the possible adjoint colour charges in this setting are $0$ and $\pm 1$. 

If the incoming gluon carries charge $e=0$ then it is aligned with the background and there is no scattering since the situation is equivalent to pure Maxwell theory. So without loss of generality, we choose the incoming gluon to have charge $e=+1$. By inspection of charge conservation at each of the cubic vertices in the ghost and gluon loops from Fig.~\ref{FIG:LOOPS}, only the colour charges $e_l=-1,0$ are allowed to run in the loop. Furthermore, a simple change of variables shows that the contributions from both of these loop charges are in fact \emph{equal}. Therefore, the minus to plus helicity flip amplitude for $N=2$ can be reduced to:
\be\begin{split}\label{su22}
  M^{\text{YM}}_{-\rightarrow +}(k)=-\frac{4\,\im\,g^2}{(2\pi)^2}  \int\limits_{-\infty}^{+\infty}\!\ud\phi \!\int\limits_0^\infty\!\ud\theta\,\theta\!\int\limits_0^1\!\ud s\,s^2 
  \,\exp\!\bigg[- \im \frac{\theta\,\text{var}(a)\,s}{k_\LCp\,(1-s)}\bigg] \,\left( a_\theta a_\theta  -\frac{1}{4}a_\phi  a_\phi \right)\,.
   \end{split}
\ee
Now define
\be\label{integra1}
\mathcal{J}_0(x) := \int\limits_0^1\!\ud s \exp\bigg[ \frac{-\im\, x\, s}{1-s}\bigg]  = 1 - \im\, x\, \e^{\im\,x} \mathrm{E}_1(\im x) \;,
\ee
where $\mathrm{E}_1$ is the exponential integral, and let
\be\label{int2}
\mathcal{J}_\text{YM}(x) := \int\limits_0^1\!\ud s\, s^2 \exp\bigg[ \frac{-\im\, x\, s}{1-s}\bigg] = \frac{1}{6}\, \partial_x^2 \big( x^2 \mathcal{J}_0(x) \big) \;.
\ee
With this notation, the helicity flip amplitude for pure Yang-Mills with two colours can be written as:
\be\label{su2f}
	M^{\text{YM}}_{-\rightarrow +}(k)=-\frac{4\,\im\,g^2}{(2\pi)^2}  \int\limits_{-\infty}^{+\infty}\!\ud\phi \!\int\limits_0^\infty\!\ud\theta\,\theta  \left( a_\theta a_\theta  -\frac{1}{4}a_\phi  a_\phi \right)\,\mathcal{J}_\text{YM}\!\left(\frac{\theta\,\mathrm{var}(a)}{k_\LCp}\right)\,,
\ee
with only the lightfront integrals remaining.

In QED, the integral $\mathcal{J}_0$ appears in \emph{tree-level} photon emission amplitudes in plane waves~\cite{Dinu:2013hsd}, while fermion loops lead to Bessel functions~\cite{Dinu:2013gaa}. The presence of trigonometric integrals rather than Bessel functions in \eqref{su2f} is a non-abelian effect. Heuristically, exponents going like $1/s(1-s)$ lead to Bessel functions, but the dressing of the gluons introduces $(e_l+se)^2$ factors in the numerator, which leads to the trigonometric integrals seen here and in QED at tree level.

\medskip

To obtain the QCD result, we must evaluate the quark loop contribution to \eqref{QCDhf2} with two colours. In this case, the two fundamental vectors of SU$(2)$ have weights $\pm\frac{1}{2}$ as
\be\label{fundweight}
\mathsf{T}^0 \begin{pmatrix} 1 \\ 0 \end{pmatrix} = \frac{1}{2}  \begin{pmatrix} 1 \\ 0 \end{pmatrix}\,, \qquad \mathsf{T}^0 \begin{pmatrix} 0 \\ 1 \end{pmatrix} = -\frac{1}{2}  \begin{pmatrix} 0 \\ 1 \end{pmatrix} \;.
\ee
Charge conservation dictates that only the fundamental weight $\mu_l=-\frac{1}{2}$ can flow in the quark loop in Fig. \ref{FIG:LOOPS}. Thus, the quark contribution to helicity flip in SU(2) is
\be\begin{split}\label{qsu21}
   \frac{\im\,g^2}{(2\pi)^2} \sum_{i=1}^{n_f} \int\limits_{-\infty}^{+\infty}\!\ud\phi \!\int\limits_0^\infty\!\ud\theta\,\theta\!\int\limits_0^1\!\ud s 
  \,\exp\!\bigg[\frac{-\im\,\theta\,\mathrm{M}_i^2[-\tfrac12]}{2s(1-s)k_\LCp} \bigg] \,\left(\frac{1}{2}- s\right)^2  \left( a_\theta a_\theta  -\frac{1}{4}a_\phi  a_\phi \right)\,. 
   \end{split}
\ee
The $s$-integral here can be performed in the case of \textit{massless} quarks. Then the relevant integral is
\be\label{JQCD}
	\mathcal{J}_\text{QCD}(x):= \int\limits_0^\infty\!\ud s\, \Big(\frac{1}{2} -s\Big)^2 \exp \bigg[\frac{-\im x(1/2-s)^2}{s(1-s)} \bigg] = \frac{\sqrt{\pi}}{16} U\Big(\frac32, 0, \im x\Big) \;,
\ee
where $U$, the confluent hypergeometric function, may be written as a sum of two linear functions multiplying Bessel functions of the zeroth and first order. It is interesting to note that the Bessels appearing in the QED helicity flip amplitude are of the same order.

Combining the Yang-Mills and quark contributions \eqref{su2f} and \eqref{qsu21}, and using \eqref{JQCD}, gives the full helicity flip amplitude for SU(2) QCD with $n_f$ massless quarks as
\be\label{SU2final}
	M^{\text{QCD}}_{-\rightarrow +}(k)=\frac{-\im\,g^2}{(2\pi)^2}  \int\limits_{-\infty}^{+\infty}\!\ud\phi \!\int\limits_0^\infty\!\ud\theta\,\theta  \bigg( a_\theta a_\theta  -\frac{1}{4}a_\phi  a_\phi \bigg)\,\bigg[ 4 \mathcal{J}_\text{YM}\!\left(\frac{\theta\,\mathrm{var}(a)}{k_\LCp}\right)- n_f \mathcal{J}_\text{QCD}\!\bigg(\frac{\theta\,\mathrm{var}(a)}{k_\LCp}\bigg)\bigg]\,,
\ee
The remaining lightfront time integrals can only be performed analytically in special cases. One example is a constant plane wave $F_{\mu\nu}$, but as this is not a `sandwich' wave it does not admit the required scattering boundary conditions. 

\begin{figure}[t!!]
		\centering\includegraphics[width=0.5\textwidth]{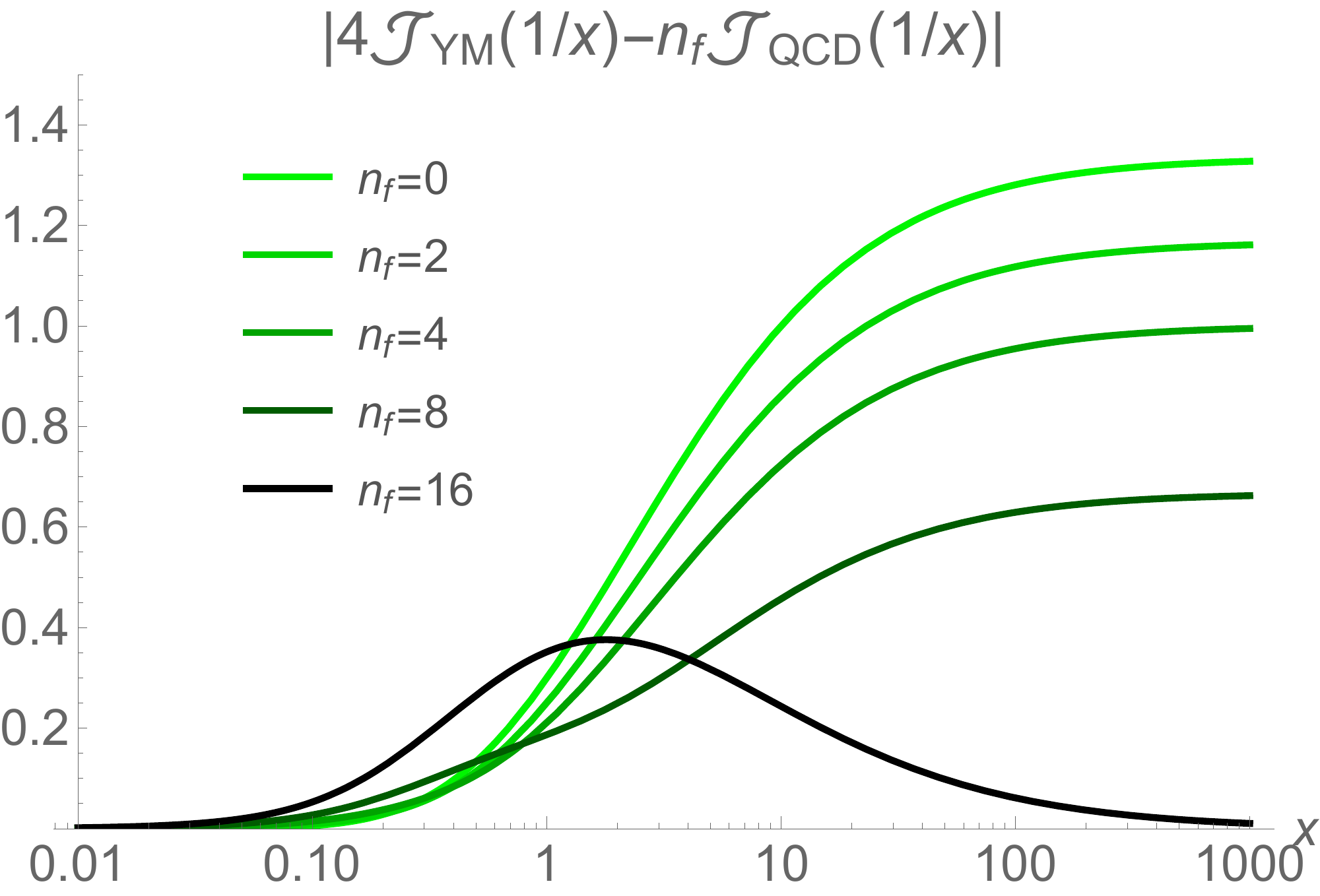}
\caption{\label{FIG:SKILLNAD} Absolute value of $\mathcal{Q}(x)$ as defined in (\ref{Funktionen}), which results from performing all loop momentum integrals in SU(2) QCD, as a function of argument $x$ and for different numbers of fundamental flavours. For $n_f=16$ the function vanishes as $x\to\infty$, which implies a suppression of the helicity-flip probability in the high-energy limit.}
\end{figure}

We therefore proceed to investigate some limits, which also apply for general numbers of colours and flavours. As a motivation for this, and in the interests of providing some more particular information about the SU(2) case here, consider the function 
\be\label{Funktionen}
	\mathcal{Q}(x) := 4 \mathcal{J}_\text{YM}\bigg(\frac{1}{x}\bigg) - n_f \mathcal{J}_\text{QCD}\!\bigg(\frac1x\bigg) \;,
\ee
defined by the combination in square brackets of the SU(2) result (\ref{SU2final}). This function is plotted in Fig.~\ref{FIG:SKILLNAD} for various numbers of flavours $n_f$. As this number increases, we notice a change in the asymptotic behaviour of the function. In particular, for $n_f=16$ the function quickly vanishes for large argument. Indeed, it can be verified using the explicit representation of $\mathcal{Q}$ in terms of special functions that
\be\label{Qf}
	\mathcal{Q}(\infty) = \frac{4}{3} - \frac{n_f}{12} \;,
\ee
vanishing at $n_f=16$. We will show below that this behaviour corresponds to a `flavour-suppression' of the helicity flip amplitude at high energy.

%%%%%%%%%%%%%%%%%%%%%%%%%%%%%%%%%%

\subsection{High lightfront energy limit}
The helicity-flip probability depends on the gluon momentum only through the lightfront component $k_\LCp$. Hence it is natural to consider how the probability behaves in the limiting cases of large and small\footnote{Strictly we should take the limit of a dimensionless invariant, for example $b = \omega_0 n\cdot k / m^2 \equiv \omega_0 k_\LCp /m^2$ for $\omega_0$ the typical (lightfront) energy scale of the background, set by its duration or frequency of oscillation, for example. The final conclusions are the same if we work simply with $k_\LCp$, however.} $k_\LCp$.

%We start with the quark loop contributions~\eqref{QCDhf2}. 

\begin{figure}[t!]
	\centering\includegraphics[width=0.5\textwidth]{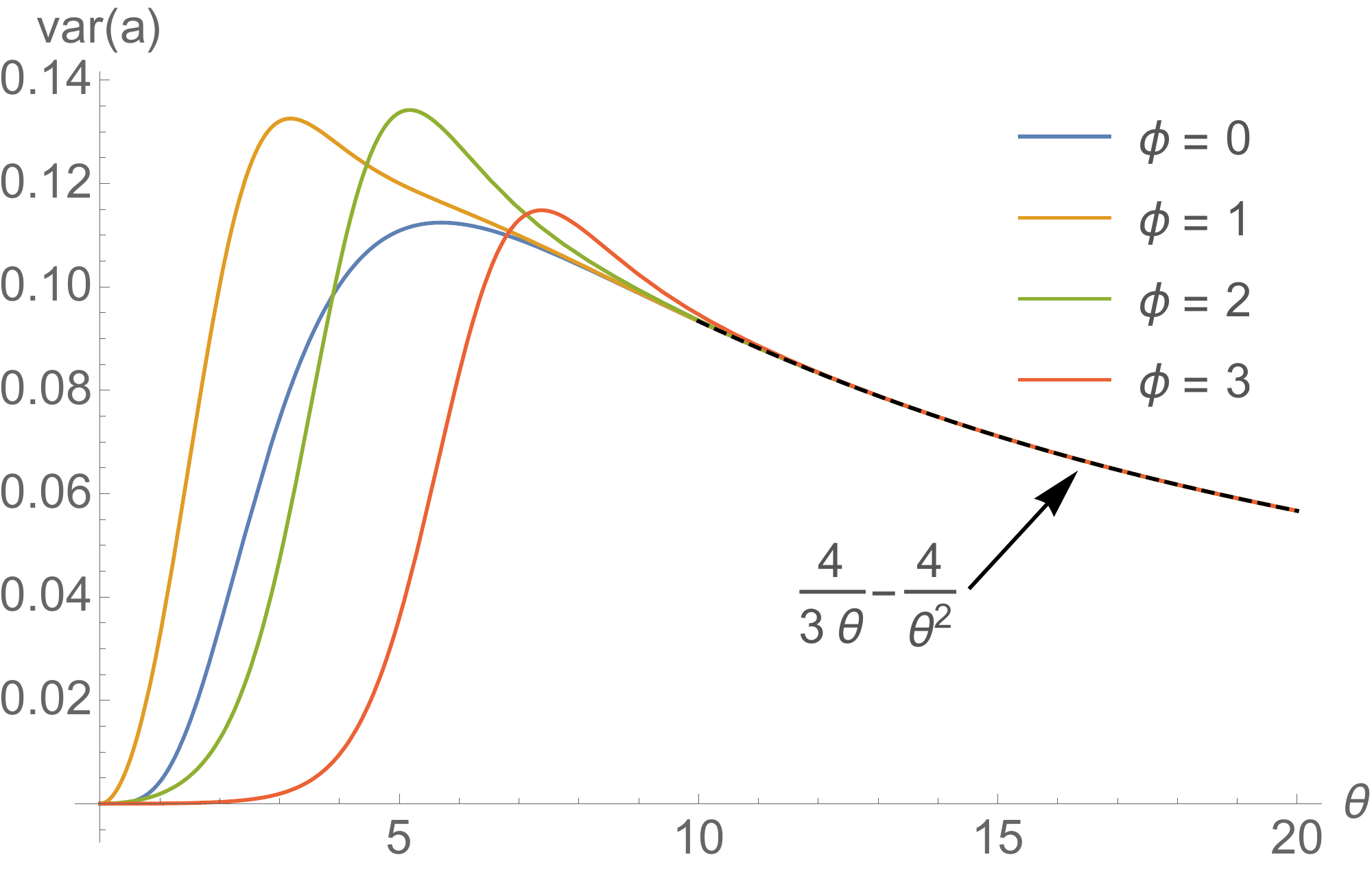}
	\caption{\label{FIG:VARPLOT} The variance in a background $a(x^\LCm) = \text{sech}^2(\omega x^\LCm)$, as a function of dimensionless variables $\phi$ and $\theta$ of the form $\omega x^\LCm$. The behaviour shown is typical of all sandwich plane waves; as a function of $\theta$, the variance rises from zero to some peak value, and then falls asymptotically back to zero as (to leading order) $1/\theta$ with $\phi$-independent coefficient~\cite{Hebenstreit:2010cc}. This asymptotic behaviour is highlighted on the plot.}
\end{figure}

Given either the YM (\ref{MYM2}) or quark \eqref{QCDhf2} loop contributions for general $N$ and $n_f$, we begin by rescaling $\theta \to \theta k_\LCp$. The exponentials appearing then depends on $k_\LCp$ only through the variance, $\text{var}({a}) \equiv \text{var}(a) (\phi,k_\LCp\theta)$, a function of $\phi$ and $k_{\LCp}\theta$. For large $k_\LCp$ we may approximate this by $\text{var}(a) (\phi, \infty)$ which, crucially, vanishes for any sandwich plane wave\footnote{Note that the variance appears as a `correction' to the mass terms in the exponentials of the loops, for both the quarks and gluons. For a long sandwich wave comprising many regular oscillations, the variance can become approximately constant over a large portion of $\phi-\theta$ space, behaving as, for e.g.~circular polarization: $m^2+2\text{var}(\hat{a}) \sim m^2(1+a_0^2) - m^2 a_0^2\text{sinc}^2(\theta/2) $~\cite{Kibble:1975vz}. The combination $m^2(1+a_0^2)$ is referred to as the electron `mass shift' in the QED literature. The nature, and observability, of this mass shift was for a long time a debated concept. It is \emph{not} a shift in the particle rest mass: no states in the theory have mass-squared equal to $m^2(1+a_0^2)$~\cite{Ilderton:2012qe}. Instead the shift (or rather the variance) encodes nonlinearities which have an observable impact on particle spectra, and which can be understood as resonance phenomena; this is made clear by considering a (sandwich) wave train of a finite number of cycles~\cite{Heinzl:2010vg}, whereas historical controversies resulted from the consideration of only infinite-duration, rather than sandwich, plane waves. (For historical references see~\cite{Ilderton:2012qe,Harvey:2012ie}.)} with $1/k_\LCp$~corrections, as~\cite{Hebenstreit:2010cc}
\be
	\text{var}({a})(\phi,\theta) \sim \frac{1}{\theta}\int\limits_{-\infty}^\infty\!\ud \phi\, |a(\phi)|^2 - \frac{1}{\theta^2}\bigg| \int\limits_{-\infty}^\infty\!\ud \phi\,  a(\phi)\bigg|^2 \quad \text{as}\quad \theta\to\infty
\ee
The typical behaviour of the variance is shown in Fig.~\ref{FIG:VARPLOT}. 
With this, the exponentials reduce to \textit{unity} for YM loops and to 
\be\label{enkelexp}
	\exp\bigg[\frac{-\im\, \theta}{k_\LCp s(1-s)} \frac{m_i^2}{2}\bigg]
\ee
for quark loops. In either case, the integrand of $M(k)$ then depends on $\phi$ only through the terms outside the exponential. In order to extract the leading order behaviour of these terms as $k_\LCp\to\infty$, it is helpful to Fourier transform, using~\eqref{urgh},
\be\begin{split}
	a(\phi+ k_\LCp \theta/2) - \langle a \rangle = \int\!\frac{\ud \omega}{2\pi}\, {a}(\omega)\, e^{\im\omega \phi} \big[ e^{\im \omega k_\LCp\theta/2}- \text{sinc}(\omega k_\LCp\theta/2)\big] \;.
\end{split}
\ee
and similarly for $a(\phi- k_\LCp \theta/2) - \langle a \rangle$. Performing the $\phi$-integral sets the Fourier variables appearing to be equal and opposite. The $\theta$-integral is then a combination of powers and trigonometric functions which can be performed, with the result (ignoring purely numerical factors):
\be\label{quark-b-high}
%	\frac{M(k)}{k_\LCp} \sim \frac{1}{b} \int\limits_0^1\!\ud s\!\int\! \frac{\ud \omega}{2\pi} \, \hat{a}(\omega)\hat{a}(-\omega)  \bigg[\frac{1}{2} + \mathcal{O}\bigg(\frac{\log b}{b^2}\bigg)\bigg] \;. 
	\frac{M(k)}{k_\LCp} \sim \frac{1}{k_\LCp} \int\limits_0^1\!\ud s\!\int\! \frac{\ud \omega}{2\pi} \, (\text{charge})^2\, a(\omega)a(-\omega)  \bigg[\frac{1}{2} + \mathcal{O}\bigg(\frac{\log b}{b^2}\bigg)\bigg] \;,
\ee
where $b := k_\LCp \omega /m^2$ and `charge' stands for the appropriate adjoint or fundamental charge. The leading order term, as expected of a high energy limit, is independent of the mass and applies to both the quark and YM  loops. (For the latter, the subleading terms in (\ref{quark-b-high}) are identically zero.) The remaining integral over $s$ can then be performed exactly.

It follows that the helicity flip probability behaves, in the limit $k_\LCp \to \infty$, as $\mathbb{P} \sim (a_0^2/k_\LCp)^2$ where $a_0$ typifies the background field strength. This is the same scaling as would be obtained if the background had been treated to leading order in perturbation theory, rather than exactly as here. The reason for this is that higher order terms in $a$, encoded in the exponential terms of the the integrands above, vanish when the variance of the field is replaced by its asymptotic limit.

The same dependence on field strength and intensity is seen in QED~\cite{Podszus:2018hnz,Ilderton:2019kqp}. However, here we have an additional dependence on the number of colours and flavours, which plays a role.  While the high-energy dependence on momentum and field strength in the YM and quark loops is the same, they differ in overall factors and in the $s$-integrations, which are affected by the charge assignments. These factors can conspire to \textit{suppress} the general leading-order behaviour identified here. Consider again SU(2) and the negative-to-positive flip with $e=1$. Then we have from \eqref{quark-b-high} the relative YM and quark contributions

\be
	2\!\sum\limits_{e_l = 0,-1} \int\limits_0^1\!\ud s\, (e_l +s)^2 = \frac{4}{3} \;, \qquad -n_f\!\sum\limits_{\mu = -\frac12} \int\limits_0^1\!\ud s\, (\mu +s)^2 = -\frac{n_f}{12} \;,
\ee
respectively, where the leading $2$ is the number of colours. Hence, for $n_f=16$ the leading order contribution to helicity flip \textit{vanishes} in the high energy limit, with the YM contribution being cancelled by that from the large number of flavours. This recovers the behaviour seen in the exact SU(2) expressions \eqref{SU2final} -- \eqref{Qf}, demonstrating that behaviour indeed corresponds to a suppression of the flip probability at high energy. This effect is clearly related to the influence of the number of flavours on the running coupling and asymptotic freedom, both because of the diagrams considered and the relative minus sign in the contributions of gluons and fermions. The reason that the critical number of fermions does not match that at which the beta-function changes sign ($n_f=11$ for $N=2$) is that helicity flip comes from the \textit{finite} parts of the diagrams considered, not the divergent parts. Related, of course, is the fact that the tadpole gives no contribution here.

%%%%%%%%%%%%%%%%%%%%%%%%%

\subsection{Low lightfront energy limit}
\label{SECT:IR}

Now consider the low (lightfront) energy limit, $k_\LCp\to 0$; this probes the IR structure of the helicity flip observable. This limit is taken by again rescaling $\theta \to k_\LCp\theta $ and expanding in powers of~$k_\LCp$. For the quark loop, the exponent immediately reduces again to \eqref{enkelexp}. This allows (expanding also the pre-exponential terms) the $\theta$-integral to be made convergent by a contour rotation (or equivalently by adding $-\im\epsilon$ to the mass), to arrive at the lowest order contribution, ignoring prefactors,
\be\begin{split}\label{quark-b-2}
	\frac{M(k)}{k_\LCp} &\sim  k_\LCp \!\int\limits_0^1\!\ud s\, s^2(1-s)^2\! \int\limits_{-\infty}^{+\infty}\!\ud\phi\,  (\mu_{l}+ e s)^2 a'(\phi)^2  \;.
\end{split}
\ee
The remaining $s$-integral can be performed directly once the charges are known. Let $a_0$ again characterise the background field strength; then the essential scaling of the flip probability at small $k_\LCp$ is $\mathbb{P} \sim (k_\LCp a_0^2)^2$, which increases from zero with increasing gluon energy $k_\LCp$. 

Things are different for the YM loops, however: these diverge as $k_\LCp\to 0$. The reason is that there is no mass term in the YM exponents: the expansion of the YM exponential terms in powers of $\theta$ begins at order $\theta^3$ because the gluon is massless. The presence of the exponential is needed for the integrals to converge, and so to find the leading order low-energy behaviour we change variables $\theta \to k_\LCp^{1/3}\theta $ and expand, such that, again ignoring purely numerical factors,
\be\begin{split}\label{glue-med-b}
	\frac{M(k)}{k_\LCp} \sim \frac{1}{k^{1/3}_\LCp} \!\int\limits_0^1\!\ud s\! \int\limits_{-\infty}^{+\infty}\!\ud\phi \int\limits_0^\infty\! \ud\theta\, \theta\, (e_l+e\, s)^2 a'(\phi)^2 \,   \exp\bigg[- \im  (e_l+e\, s)^2\frac{\theta^3 a'(\phi) \bar{a}'(\phi)}{12s(1-s)} \bigg] +\cdots 
\end{split}
\ee
Performing the $\theta$ integral at leading order gives
\be\begin{split}\label{glue-med-b2}
	\frac{M(k)}{k_\LCp} \sim \frac{1}{k_\LCp^{1/3}}\!\int\limits_0^1\!\ud s  (s(1-s))^{2/3} \! \int\!\ud\phi\, \frac{\big[(e_l+e\, s)a'(\phi)\big]^{4/3}}{ \big[(e_l+e\, s){\bar a}'(\phi)\big]^{2/3}}   \;.
\end{split}
\ee
Note that there are no divergences in the $\phi$ or $s$ integrals, and that a dependence on the anti-holomorphic part of the background has re-appeared through contributions from the effective mass. The probability then scales as $\mathbb{P} \sim \big(a_0^{2/3} / k_\LCp^{1/3}\big)^2$ and so is divergent as $k_\LCp\to 0$. This limit encompasses both the soft limit ($k_0 \to 0$) as well as the collinear limit, in which the gluon momentum is parallel to the background direction, $k_\mu \to \lambda  n_\mu$ for some $\lambda$. The divergence is infra-red, as it results from the large-distance divergence of the $\theta$-integral when $k_\LCp = 0$.

%The Bessel functions reappear in the quark loop. \Anton{Specialising to massless quarks for simplicity,}  the relevant integral is
%
%\be
%	\mathcal{J}_\text{QCD}(x):= \int\limits_0^\infty\!\ud s\, \Big(1-\frac{s}{2}\Big)^2 \exp \bigg[\frac{-i x(1-s/2)^2}{s(1-s)} \bigg] = \frac{\sqrt{\pi}}{16} U\Big(\frac32, 0, i x\Big) \;,
%\ee
%
%where $U$, the confluent hypergeometric function, may be written as a sum of two linear functions multiplying Bessel functions of the zeroth and first order. Bessels of the same order appear in the QED helicity flip amplitude. 
%
%With this the helicity-flip amplitude, and by extension the probability (\ref{}), is reduced to a double integral over two lightfront times which may be evaluated numerically.  
%

%%%%%%%%%%%%%%%%%%%%%%%%%%%
%%%%%%%%%%%%%%%%%%%%%%%%%%%

\section{Conclusions}
\label{Conc}
 
In this paper, we have computed the leading contribution to gluon helicity flip in a gauge theory plane wave background. Since the Feynman rules for perturbative Yang-Mills and QCD in the plane wave background can be determined exactly, this calculation is performed without recourse to approximations: the plane wave background is treated exactly. Using pure helicity polarization states, we were able to obtain compact formulae for the helicity flip amplitude, \eqref{MYM2} and \eqref{QCDhf2}, for any number of colours or flavours. These expressions are further simplified by making an explicit choice for the number of colours, or by taking the high- or low-lightfront energy limit of the probe gluon.

While in some ways analogous to the calculation of photon helicity flip in strong field QED with a plane wave background~\cite{Dinu:2013gaa}, there are substantial differences in the non-abelian theory. Here, all particles are dressed by the background, leading to different functional dependencies in the amplitude. For instance, the exponential factors appearing in the Yang-Mills and QCD helicity flip amplitudes show more structure than their QED `Volkov exponent' counterparts. We showed that this difference leads to the appearance of trigonometric and confluent hypergeometric functions, as opposed to the Bessel functions that appear in QED at one loop. Furthermore, by tuning the number of flavours, we observed a `flavour suppression' of the helicity flip amplitude in the high lightfront energy limit. In the low lightfront energy limit, we found soft and collinear divergences.

These are non-abelian phenomena, and one could wonder if flavour suppression is related to a conformal fixed point of QCD, where the 1-loop beta function changes sign. Yet it is easy to see that this occurs at a different value of $n_f$ then flavour suppression (e.g., for $N=2$ the 1-loop beta function vanishes for $n_f=11$ rather than $16$). This is not surprising, as the flat background diagrams (including the entire tadpole term) do not contribute to helicity flip. However, it would be interesting to know if there is a physical explanation for flavour suppression beyond the fact that quarks contribute to the amplitude with the opposite sign to the pure Yang-Mills components.

\medskip

To conclude, we outline two potential applications of our results. The first is in the context of formal quantum field theory, pertaining to a structure known as \emph{double copy}. This is a prescription which relates scattering amplitudes of gauge theory and gravity, perturbatively around a trivial scattering background (cf., \cite{Kawai:1985xq,Bern:2008qj,Bern:2010ue,Bern:2010yg}). The computational power of double copy has now been demonstrated in a variety of scenarios: from high-loop calculations in a variety of supergravity theories~\cite{Bern:2013uka,Bern:2014sna,Bern:2017ucb,Bern:2018jmv}, to perturbative determination of far-field gravitational radiation~\cite{Goldberger:2016iau,Luna:2017dtq,Goldberger:2017ogt,Shen:2018ebu}, to state-of-the-art calculations of the conservative two-body Hamiltonian in the post-Minkowskian expansion of general relativity~\cite{Cheung:2018wkq,Bern:2019nnu}. Nevertheless, the origin and robustness of double copy remains mysterious: it lacks a fully non-linear Lagrangian or off-shell explanation at the level of the underlying gauge and gravitational field theories.

One way to probe the robustness of double copy is by testing the extent to which it holds for scattering amplitudes on non-trivial perturbative backgrounds~\cite{Adamo:2017nia,Bahjat-Abbas:2017htu,Carrillo-Gonzalez:2017iyj}, such as a plane wave. At lowest order double copy on a plane wave background has been verified~\cite{Adamo:2017nia}, but to make more general statements requires generating gauge theory data to feed into any double copy prescription. For instance, the $2\to 2$ gluon tree amplitude on a plane wave background was only recently obtained in a form suitable to double copy~\cite{Adamo:2018mpq}. 

The gluon helicity flip amplitude can be viewed as data which can be used to test double copy on a plane wave background \emph{beyond} tree-level. Indeed, we conjecture that the results of this paper can be used to determine the \emph{graviton} helicity flip amplitude on a plane wave metric background. To test this requires computing the graviton helicity flip amplitude directly from the Einstein-Hilbert action; we hope to pursue this calculation in the future.

\medskip

A second potential application is to the colour glass condensate (CGC), the effective theory describing strong gauge fields in the vicinity of heavy ion collisions~\cite{Iancu:2002xk,Iancu:2003xm,Gelis:2010nm,Kovchegov:2012mbw,Blaizot:2016qgz}. As mentioned in the Introduction, the gauge fields of the CGC are strong and are thus treated exactly, as non-perturbative background fields. The classical $A_\mu$ describing the CGC is characterised by purely transverse fields, which are orthogonal and of equal magnitude: $\vec{E}\cdot \vec{B}=\vec{E}^2-\vec{B}^2=0$, for $\vec{E}$, $\vec{B}$ the electric and magnetic fields of $A_\mu$ (for each colour)~\cite{Iancu:2002xk}.

These gauge fields are represented by potentials of the form $\alpha(x^\LCperp,x^\LCm) \ud x^\LCm$, where $-\nabla_\LCperp^2\alpha = \rho$, with $\rho$ a colour source (localised around $x^\LCm=0$). The plane waves considered in this paper are source-free members of this class, with $\alpha=x^{\LCperp}\dot{a}_{\perp}(x^\LCm)$, and obey the same relations between $\vec E$ and $\vec B$ as above. Calculations involving the CGC require averaging over the classical fields $A$ from sources distributed with a Gaussian weight~\cite{Iancu:2002xk} around $\rho=0$. It could be useful to replace the averaging with a `dominant contribution' from the peak of the Gaussian. If so, this dominant contribution could, following the observations above, be a plane wave. This will be addressed in more detail elsewhere.

\begin{acknowledgments}
We thank Eduardo Casali, Tom Heinzl and Piotr Tourkine for many useful discussions and comments on the manuscript. TA is supported by an Imperial College Junior Research Fellowship.
\end{acknowledgments}

\bibliographystyle{JHEP}
%\bibliography{GluePlaneBib}

\begin{thebibliography}{10}

\bibitem{Jaeckel:2010ni}
J.~Jaeckel and A.~Ringwald, \emph{{The Low-Energy Frontier of Particle
  Physics}},
  \href{https://doi.org/10.1146/annurev.nucl.012809.104433}{\emph{Ann. Rev.
  Nucl. Part. Sci.} {\bfseries 60} (2010) 405}
  [\href{https://arxiv.org/abs/1002.0329}{{\ttfamily 1002.0329}}].

\bibitem{Dobrich:2010hi}
B.~Dobrich and H.~Gies, \emph{{Axion-like-particle search with high-intensity
  lasers}}, \href{https://doi.org/10.1007/JHEP10(2010)022}{\emph{JHEP}
  {\bfseries 10} (2010) 022} [\href{https://arxiv.org/abs/1006.5579}{{\ttfamily
  1006.5579}}].

\bibitem{Redondo:2010dp}
J.~Redondo and A.~Ringwald, \emph{{Light shining through walls}},
  \href{https://doi.org/10.1080/00107514.2011.563516}{\emph{Contemp. Phys.}
  {\bfseries 52} (2011) 211} [\href{https://arxiv.org/abs/1011.3741}{{\ttfamily
  1011.3741}}].

\bibitem{Dunne:2008kc}
G.~V. Dunne, \emph{{New Strong-Field QED Effects at ELI: Nonperturbative Vacuum
  Pair Production}},
  \href{https://doi.org/10.1140/epjd/e2009-00022-0}{\emph{Eur. Phys. J.}
  {\bfseries D55} (2009) 327}
  [\href{https://arxiv.org/abs/0812.3163}{{\ttfamily 0812.3163}}].

\bibitem{DeWitt:1967ub}
B.~S. DeWitt, \emph{{Quantum Theory of Gravity. 2. The Manifestly Covariant
  Theory}}, \href{https://doi.org/10.1103/PhysRev.162.1195}{\emph{Phys. Rev.}
  {\bfseries 162} (1967) 1195}.

\bibitem{tHooft:1975uxh}
G.~'t~Hooft, \emph{{The Background Field Method in Gauge Field Theories}},  in
  \emph{{Functional and Probabilistic Methods in Quantum Field Theory. 1.
  Proceedings, 12th Winter School of Theoretical Physics, Karpacz, Feb 17-March
  2, 1975}}, pp.~345--369, 1975.

\bibitem{Boulware:1980av}
D.~G. Boulware, \emph{{Gauge Dependence of the Effective Action}},
  \href{https://doi.org/10.1103/PhysRevD.23.389}{\emph{Phys. Rev.} {\bfseries
  D23} (1981) 389}.

\bibitem{Abbott:1981ke}
L.~F. Abbott, \emph{{Introduction to the Background Field Method}}, {\emph{Acta
  Phys. Polon.} {\bfseries B13} (1982) 33}.

\bibitem{Furry:1951zz}
W.~H. Furry, \emph{{On Bound States and Scattering in Positron Theory}},
  \href{https://doi.org/10.1103/PhysRev.81.915}{\emph{Phys. Rev.} {\bfseries
  81} (1951) 115}.

\bibitem{Wolkow:1935zz}
D.~M. Wolkow, \emph{{\"Uber eine Klasse von Losungen der Diracschen
  Gleichung}}, \href{https://doi.org/10.1007/BF01331022}{\emph{Z. Phys.}
  {\bfseries 94} (1935) 250}.

\bibitem{Seipt:2017ckc}
D.~Seipt, \emph{{Volkov States and Non-linear Compton Scattering in Short and
  Intense Laser Pulses}},  in \emph{{Proceedings, QFT at the Limits: from
  Strong Fields to Heavy Quarks (2016) Dubna, Russia}}, pp.~24--43, 2017,
  \href{https://arxiv.org/abs/1701.03692}{{\ttfamily 1701.03692}}.

\bibitem{RitusRev}
V.~I. Ritus, \emph{Quantum effects of the interaction of elementary particles
  with an intense electromagnetic field}, {\emph{J. Russ. Laser Res.}
  {\bfseries 6} (1985) 497}.

\bibitem{DiPiazza:2011tq}
A.~Di~Piazza, C.~Muller, K.~Z. Hatsagortsyan and C.~H. Keitel, \emph{{Extremely
  high-intensity laser interactions with fundamental quantum systems}},
  \href{https://doi.org/10.1103/RevModPhys.84.1177}{\emph{Rev. Mod. Phys.}
  {\bfseries 84} (2012) 1177}
  [\href{https://arxiv.org/abs/1111.3886}{{\ttfamily 1111.3886}}].

\bibitem{King:2015tba}
B.~King and T.~Heinzl, \emph{{Measuring Vacuum Polarisation with High Power
  Lasers}},  \href{https://arxiv.org/abs/1510.08456}{{\ttfamily 1510.08456}}.

\bibitem{Adamo:2018mpq}
T.~Adamo, E.~Casali, L.~Mason and S.~Nekovar, \emph{{Plane wave backgrounds and
  colour-kinematics duality}},
  \href{https://arxiv.org/abs/1810.05115}{{\ttfamily 1810.05115}}.

\bibitem{Basler:1984hu}
M.~Basler and A.~Hadicke, \emph{{On nonabelian SU(2) plane waves}},
  \href{https://doi.org/10.1016/0370-2693(84)90180-1}{\emph{Phys. Lett.}
  {\bfseries 144B} (1984) 83}.

\bibitem{Adamo:2017nia}
T.~Adamo, E.~Casali, L.~Mason and S.~Nekovar, \emph{{Scattering on plane waves
  and the double copy}},
  \href{https://doi.org/10.1088/1361-6382/aa9961}{\emph{Class. Quant. Grav.}
  {\bfseries 35} (2018) 015004}
  [\href{https://arxiv.org/abs/1706.08925}{{\ttfamily 1706.08925}}].

\bibitem{Srednicki:2007qs}
M.~Srednicki, \emph{{Quantum field theory}}. Cambridge University Press, 2007.

\bibitem{Elvang:2013cua}
H.~Elvang and Y.-t. Huang, \emph{{Scattering Amplitudes}},
  \href{https://arxiv.org/abs/1308.1697}{{\ttfamily 1308.1697}}.

\bibitem{Dixon:2013uaa}
L.~J. Dixon, \emph{{A brief introduction to modern amplitude methods}},  in
  \emph{{Proceedings, 2012 European School of High-Energy Physics (ESHEP 2012):
  La Pommeraye, Anjou, France, June 06-19, 2012}}, pp.~31--67, 2014,
  \href{https://arxiv.org/abs/1310.5353}{{\ttfamily 1310.5353}},
  \href{https://doi.org/10.5170/CERN-2014-008.31}{DOI}.

\bibitem{Cheung:2017pzi}
C.~Cheung, \emph{{TASI Lectures on Scattering Amplitudes}},  in
  \emph{{Proceedings, Theoretical Advanced Study Institute in Elementary
  Particle Physics : Anticipating the Next Discoveries in Particle Physics
  (TASI 2016): Boulder, CO, USA, June 6-July 1, 2016}}, pp.~571--623, 2018,
  \href{https://arxiv.org/abs/1708.03872}{{\ttfamily 1708.03872}},
  \href{https://doi.org/10.1142/9789813233348_0008}{DOI}.

\bibitem{Toll:1952rq}
J.~S. Toll, \emph{{The Dispersion relation for light and its application to
  problems involving electron pairs}}, Ph.D. thesis, Princeton U., 1952.

\bibitem{Narozhny:1969}
N.~B. Narozhny, \emph{{Propagation of plane electomagnetic waves in a constant
  field}}, {\emph{JETP} {\bfseries 28} (1969) 371}.

\bibitem{Ritus:1972ky}
V.~I. Ritus, \emph{{Radiative corrections in quantum electrodynamics with
  intense field and their analytical properties}},
  \href{https://doi.org/10.1016/0003-4916(72)90191-1}{\emph{Annals Phys.}
  {\bfseries 69} (1972) 555}.

\bibitem{Shore:2007um}
G.~M. Shore, \emph{{Superluminality and UV completion}},
  \href{https://doi.org/10.1016/j.nuclphysb.2007.03.034}{\emph{Nucl. Phys.}
  {\bfseries B778} (2007) 219}
  [\href{https://arxiv.org/abs/hep-th/0701185}{{\ttfamily hep-th/0701185}}].

\bibitem{Becker:1974en}
W.~Becker and H.~Mitter, \emph{{Vacuum polarization in laser fields}},
  \href{https://doi.org/10.1088/0305-4470/8/10/017}{\emph{J. Phys.} {\bfseries
  A8} (1975) 1638}.

\bibitem{Baier:1975ff}
V.~N. Baier, A.~I. Milshtein and V.~M. Strakhovenko, \emph{{Interaction Between
  a Photon and a High Intensity Electromagnetic Wave}}, {\emph{Zh. Eksp. Teor.
  Fiz.} {\bfseries 69} (1975) 1893}.

\bibitem{Iancu:2002xk}
E.~Iancu, A.~Leonidov and L.~McLerran, \emph{{The Color glass condensate: An
  Introduction}},  in \emph{{QCD perspectives on hot and dense matter.
  Proceedings, NATO Advanced Study Institute, Summer School, Cargese, France,
  August 6-18, 2001}}, pp.~73--145, 2002,
  \href{https://arxiv.org/abs/hep-ph/0202270}{{\ttfamily hep-ph/0202270}}.

\bibitem{Iancu:2003xm}
E.~Iancu and R.~Venugopalan, \emph{{The Color glass condensate and high-energy
  scattering in QCD}},  in \emph{Quark-gluon plasma 4}, R.~C. Hwa and X.-N.
  Wang, eds., pp.~249--3363, (2003),
  \href{https://arxiv.org/abs/hep-ph/0303204}{{\ttfamily hep-ph/0303204}},
  \href{https://doi.org/10.1142/9789812795533_0005}{DOI}.

\bibitem{Gelis:2010nm}
F.~Gelis, E.~Iancu, J.~Jalilian-Marian and R.~Venugopalan, \emph{{The Color
  Glass Condensate}},
  \href{https://doi.org/10.1146/annurev.nucl.010909.083629}{\emph{Ann. Rev.
  Nucl. Part. Sci.} {\bfseries 60} (2010) 463}
  [\href{https://arxiv.org/abs/1002.0333}{{\ttfamily 1002.0333}}].

\bibitem{Kovchegov:2012mbw}
Y.~V. Kovchegov and E.~Levin, \emph{{Quantum chromodynamics at high energy}},
  vol.~33. Cambridge University Press, 2012.

\bibitem{Blaizot:2016qgz}
J.-P. Blaizot, \emph{{High gluon densities in heavy ion collisions}},
  \href{https://doi.org/10.1088/1361-6633/aa5435}{\emph{Rept. Prog. Phys.}
  {\bfseries 80} (2017) 032301}
  [\href{https://arxiv.org/abs/1607.04448}{{\ttfamily 1607.04448}}].

\bibitem{Dinu:2013gaa}
V.~Dinu, T.~Heinzl, A.~Ilderton, M.~Marklund and G.~Torgrimsson, \emph{{Vacuum
  refractive indices and helicity flip in strong-field QED}},
  \href{https://doi.org/10.1103/PhysRevD.89.125003}{\emph{Phys. Rev.}
  {\bfseries D89} (2014) 125003}
  [\href{https://arxiv.org/abs/1312.6419}{{\ttfamily 1312.6419}}].

\bibitem{Carroll}
J.-M. L\'evy-Leblond, \emph{Une nouvelle limite non-relativiste du groupe de
  poincar\'e}, {\emph{Annales de l'I.H.P. Physique th\'eorique} {\bfseries 3}
  (1965) 1}.

\bibitem{Duval:2017els}
C.~Duval, G.~W. Gibbons, P.~A. Horvathy and P.~M. Zhang, \emph{{Carroll
  symmetry of plane gravitational waves}},
  \href{https://doi.org/10.1088/1361-6382/aa7f62}{\emph{Class. Quant. Grav.}
  {\bfseries 34} (2017) 175003}
  [\href{https://arxiv.org/abs/1702.08284}{{\ttfamily 1702.08284}}].

\bibitem{Coleman:1977ps}
S.~R. Coleman, \emph{{Nonabelian Plane Waves}},
  \href{https://doi.org/10.1016/0370-2693(77)90344-6}{\emph{Phys. Lett.}
  {\bfseries 70B} (1977) 59}.

\bibitem{Kovacs:1978vb}
E.~Kovacs and S.-y. Lo, \emph{{Selfdual Propagating Wave Solutions in
  {Yang-Mills} Gauge Theory}},
  \href{https://doi.org/10.1103/PhysRevD.19.3649}{\emph{Phys. Rev.} {\bfseries
  D19} (1979) 3649}.

\bibitem{Lo:1979vq}
S.-Y. Lo, P.~Desmond and E.~Kovacs, \emph{{General selfdual nonabelian plane
  waves}}, \href{https://doi.org/10.1016/0370-2693(80)90963-6}{\emph{Phys.
  Lett.} {\bfseries 90B} (1980) 419}.

\bibitem{Trautman:1980bj}
A.~Trautman, \emph{{A class of null solutions to the Yang-Mills equations}},
  \href{https://doi.org/10.1088/0305-4470/13/1/001}{\emph{J. Phys.} {\bfseries
  A13} (1980) L1}.

\bibitem{Schwinger:1951nm}
J.~S. Schwinger, \emph{{On gauge invariance and vacuum polarization}},
  \href{https://doi.org/10.1103/PhysRev.82.664}{\emph{Phys. Rev.} {\bfseries
  82} (1951) 664}.

\bibitem{Heinzl:2017blq}
T.~Heinzl and A.~Ilderton, \emph{{Superintegrable relativistic systems in
  spacetime-dependent background fields}},
  \href{https://doi.org/10.1088/1751-8121/aa7fa3}{\emph{J. Phys.} {\bfseries
  A50} (2017) 345204} [\href{https://arxiv.org/abs/1701.09168}{{\ttfamily
  1701.09168}}].

\bibitem{Gibbons:1975jb}
G.~W. Gibbons, \emph{{Quantized Fields Propagating in Plane Wave Space-Times}},
  \href{https://doi.org/10.1007/BF01629249}{\emph{Commun. Math. Phys.}
  {\bfseries 45} (1975) 191}.

\bibitem{Deser1975}
S.~Deser, \emph{Plane waves do not polarize the vacuum},
  \href{https://doi.org/10.1088/0305-4470/8/12/012}{\emph{Journal of Physics A:
  Mathematical and General} {\bfseries 8} (1975) 1972}.

\bibitem{Dinu:2012tj}
V.~Dinu, T.~Heinzl and A.~Ilderton, \emph{{Infra-Red Divergences in Plane Wave
  Backgrounds}}, \href{https://doi.org/10.1103/PhysRevD.86.085037}{\emph{Phys.
  Rev.} {\bfseries D86} (2012) 085037}
  [\href{https://arxiv.org/abs/1206.3957}{{\ttfamily 1206.3957}}].

\bibitem{Ilderton:2012qe}
A.~Ilderton and G.~Torgrimsson, \emph{{Scattering in plane-wave backgrounds:
  infra-red effects and pole structure}},
  \href{https://doi.org/10.1103/PhysRevD.87.085040}{\emph{Phys. Rev.}
  {\bfseries D87} (2013) 085040}
  [\href{https://arxiv.org/abs/1210.6840}{{\ttfamily 1210.6840}}].

\bibitem{Heinzl:2009zd}
T.~Heinzl, A.~Ilderton and M.~Marklund, \emph{{Laser intensity effects in
  noncommutative QED}},
  \href{https://doi.org/10.1103/PhysRevD.81.051902}{\emph{Phys. Rev.}
  {\bfseries D81} (2010) 051902}
  [\href{https://arxiv.org/abs/0909.0656}{{\ttfamily 0909.0656}}].

\bibitem{VillalbaChavez:2012bb}
S.~Villalba-Chavez and C.~Muller, \emph{{Photo-production of scalar particles
  in the field of a circularly polarized laser beam}},
  \href{https://doi.org/10.1016/j.physletb.2012.11.035}{\emph{Phys. Lett.}
  {\bfseries B718} (2013) 992}
  [\href{https://arxiv.org/abs/1208.3595}{{\ttfamily 1208.3595}}].

\bibitem{Dillon:2018ypt}
B.~M. Dillon and B.~King, \emph{{ALP production through non-linear Compton
  scattering in intense fields}},
  \href{https://doi.org/10.1140/epjc/s10052-018-6207-0}{\emph{Eur. Phys. J.}
  {\bfseries C78} (2018) 775}
  [\href{https://arxiv.org/abs/1802.07498}{{\ttfamily 1802.07498}}].

\bibitem{King:2018qbq}
B.~King, \emph{{Electron-seeded ALP production and ALP decay in an oscillating
  electromagnetic field}},
  \href{https://doi.org/10.1016/j.physletb.2018.06.016}{\emph{Phys. Lett.}
  {\bfseries B782} (2018) 737}
  [\href{https://arxiv.org/abs/1802.07507}{{\ttfamily 1802.07507}}].

\bibitem{Meuren:2015iha}
S.~Meuren, C.~H. Keitel and A.~Di~Piazza, \emph{{Nonlinear neutrino-photon
  interactions inside strong laser pulses}},
  \href{https://doi.org/10.1007/JHEP06(2015)127}{\emph{JHEP} {\bfseries 06}
  (2015) 127} [\href{https://arxiv.org/abs/1504.02722}{{\ttfamily
  1504.02722}}].

\bibitem{Heinzl:2006xc}
T.~Heinzl, B.~Liesfeld, K.-U. Amthor, H.~Schwoerer, R.~Sauerbrey and A.~Wipf,
  \emph{{On the observation of vacuum birefringence}},
  \href{https://doi.org/10.1016/j.optcom.2006.06.053}{\emph{Opt. Commun.}
  {\bfseries 267} (2006) 318}
  [\href{https://arxiv.org/abs/hep-ph/0601076}{{\ttfamily hep-ph/0601076}}].

\bibitem{Karbstein:2015xra}
F.~Karbstein, H.~Gies, M.~Reuter and M.~Zepf, \emph{{Vacuum birefringence in
  strong inhomogeneous electromagnetic fields}},
  \href{https://doi.org/10.1103/PhysRevD.92.071301}{\emph{Phys. Rev.}
  {\bfseries D92} (2015) 071301}
  [\href{https://arxiv.org/abs/1507.01084}{{\ttfamily 1507.01084}}].

\bibitem{Schlenvoigt2016}
H.-P. Schlenvoigt, T.~Heinzl, U.~Schramm, T.~E. Cowan and R.~Sauerbrey,
  \emph{Detecting vacuum birefringence with x-ray free electron lasers and
  high-power optical lasers: a feasibility study},
  \href{https://doi.org/10.1088/0031-8949/91/2/023010}{\emph{Physica Scripta}
  {\bfseries 91} (2016) 023010}.

\bibitem{Mustaki:1990im}
D.~Mustaki, S.~Pinsky, J.~Shigemitsu and K.~Wilson, \emph{{Perturbative
  renormalization of null plane QED}},
  \href{https://doi.org/10.1103/PhysRevD.43.3411}{\emph{Phys. Rev.} {\bfseries
  D43} (1991) 3411}.

\bibitem{Schoonderwoerd:1998qj}
N.~C.~J. Schoonderwoerd and B.~L.~G. Bakker, \emph{{Equivalence of renormalized
  covariant and light front perturbation theory. 1. Longitudinal divergences in
  the Yukawa model}},
  \href{https://doi.org/10.1103/PhysRevD.57.4965}{\emph{Phys. Rev.} {\bfseries
  D57} (1998) 4965}.

\bibitem{Srivastava:2000cf}
P.~P. Srivastava and S.~J. Brodsky, \emph{{Light front quantized QCD in light
  cone gauge}}, \href{https://doi.org/10.1103/PhysRevD.64.045006}{\emph{Phys.
  Rev.} {\bfseries D64} (2001) 045006}
  [\href{https://arxiv.org/abs/hep-ph/0011372}{{\ttfamily hep-ph/0011372}}].

\bibitem{Mantovani:2016uxq}
L.~Mantovani, B.~Pasquini, X.~Xiong and A.~Bacchetta, \emph{{Revisiting the
  equivalence of light-front and covariant QED in the light-cone gauge}},
  \href{https://doi.org/10.1103/PhysRevD.94.116005}{\emph{Phys. Rev.}
  {\bfseries D94} (2016) 116005}
  [\href{https://arxiv.org/abs/1609.00746}{{\ttfamily 1609.00746}}].

\bibitem{Brodsky:1997de}
S.~J. Brodsky, H.-C. Pauli and S.~S. Pinsky, \emph{{Quantum chromodynamics and
  other field theories on the light cone}},
  \href{https://doi.org/10.1016/S0370-1573(97)00089-6}{\emph{Phys. Rept.}
  {\bfseries 301} (1998) 299}
  [\href{https://arxiv.org/abs/hep-ph/9705477}{{\ttfamily hep-ph/9705477}}].

\bibitem{Heinzl:2000ht}
T.~Heinzl, \emph{{Light cone quantization: Foundations and applications}},
  \href{https://doi.org/10.1007/3-540-45114-5_2}{\emph{Lect. Notes Phys.}
  {\bfseries 572} (2001) 55}
  [\href{https://arxiv.org/abs/hep-th/0008096}{{\ttfamily hep-th/0008096}}].

\bibitem{Heinzl:2003jy}
T.~Heinzl, \emph{{Light cone zero modes revisited}},  in \emph{{Light cone
  physics: Hadrons and beyond: Proceedings. 2003}}, 2003,
  \href{https://arxiv.org/abs/hep-th/0310165}{{\ttfamily hep-th/0310165}}.

\bibitem{Casher:1976ae}
A.~Casher, \emph{{Gauge Fields on the Null Plane}},
  \href{https://doi.org/10.1103/PhysRevD.14.452}{\emph{Phys. Rev.} {\bfseries
  D14} (1976) 452}.

\bibitem{Dittrich:2000zu}
W.~Dittrich and H.~Gies, \emph{{Probing the quantum vacuum. Perturbative
  effective action approach in quantum electrodynamics and its application}},
  \href{https://doi.org/10.1007/3-540-45585-X}{\emph{Springer Tracts Mod.
  Phys.} {\bfseries 166} (2000) 1}.

\bibitem{Adamo:2017qyl}
T.~Adamo, \emph{{Lectures on twistor theory}},
  \href{https://doi.org/10.22323/1.323.0003}{\emph{PoS} {\bfseries Modave2017}
  (2018) 003} [\href{https://arxiv.org/abs/1712.02196}{{\ttfamily
  1712.02196}}].

\bibitem{Witten:2003nn}
E.~Witten, \emph{{Perturbative gauge theory as a string theory in twistor
  space}}, \href{https://doi.org/10.1007/s00220-004-1187-3}{\emph{Commun. Math.
  Phys.} {\bfseries 252} (2004) 189}
  [\href{https://arxiv.org/abs/hep-th/0312171}{{\ttfamily hep-th/0312171}}].

\bibitem{Kibble:1965zza}
T.~W.~B. Kibble, \emph{{Frequency Shift in High-Intensity Compton Scattering}},
  \href{https://doi.org/10.1103/PhysRev.138.B740}{\emph{Phys. Rev.} {\bfseries
  138} (1965) B740}.

\bibitem{Kibble:1975vz}
T.~W.~B. Kibble, A.~Salam and J.~A. Strathdee, \emph{{Intensity Dependent Mass
  Shift and Symmetry Breaking}},
  \href{https://doi.org/10.1016/0550-3213(75)90581-7}{\emph{Nucl. Phys.}
  {\bfseries B96} (1975) 255}.

\bibitem{Hebenstreit:2010cc}
F.~Hebenstreit, A.~Ilderton, M.~Marklund and J.~Zamanian, \emph{{Strong field
  effects in laser pulses: the Wigner formalism}},
  \href{https://doi.org/10.1103/PhysRevD.83.065007}{\emph{Phys. Rev.}
  {\bfseries D83} (2011) 065007}
  [\href{https://arxiv.org/abs/1011.1923}{{\ttfamily 1011.1923}}].

\bibitem{Harvey:2012ie}
C.~Harvey, T.~Heinzl, A.~Ilderton and M.~Marklund, \emph{{Intensity-Dependent
  Electron Mass Shift in a Laser Field: Existence, Universality, and
  Detection}},
  \href{https://doi.org/10.1103/PhysRevLett.109.100402}{\emph{Phys. Rev. Lett.}
  {\bfseries 109} (2012) 100402}
  [\href{https://arxiv.org/abs/1203.6077}{{\ttfamily 1203.6077}}].

\bibitem{Dinu:2013hsd}
V.~Dinu, \emph{{Exact final state integrals for strong field QED}},
  \href{https://doi.org/10.1103/PhysRevA.87.052101}{\emph{Phys. Rev.}
  {\bfseries A87} (2013) 052101}
  [\href{https://arxiv.org/abs/1302.1513}{{\ttfamily 1302.1513}}].

\bibitem{Heinzl:2010vg}
T.~Heinzl, A.~Ilderton and M.~Marklund, \emph{{Finite size effects in
  stimulated laser pair production}},
  \href{https://doi.org/10.1016/j.physletb.2010.07.044}{\emph{Phys. Lett.}
  {\bfseries B692} (2010) 250}
  [\href{https://arxiv.org/abs/1002.4018}{{\ttfamily 1002.4018}}].

\bibitem{Podszus:2018hnz}
T.~Podszus and A.~Di~Piazza, \emph{{On the High-Energy Behavior of Strong-Field
  QED in an Intense Plane Wave}},
  \href{https://arxiv.org/abs/1812.08673}{{\ttfamily 1812.08673}}.

\bibitem{Ilderton:2019kqp}
A.~Ilderton, \emph{{A note on the conjectured breakdown of QED perturbation
  theory in strong fields}},
  \href{https://arxiv.org/abs/1901.00317}{{\ttfamily 1901.00317}}.

\bibitem{Kawai:1985xq}
H.~Kawai, D.~C. Lewellen and S.~H.~H. Tye, \emph{{A Relation Between Tree
  Amplitudes of Closed and Open Strings}},
  \href{https://doi.org/10.1016/0550-3213(86)90362-7}{\emph{Nucl. Phys.}
  {\bfseries B269} (1986) 1}.

\bibitem{Bern:2008qj}
Z.~Bern, J.~J.~M. Carrasco and H.~Johansson, \emph{{New Relations for
  Gauge-Theory Amplitudes}},
  \href{https://doi.org/10.1103/PhysRevD.78.085011}{\emph{Phys. Rev.}
  {\bfseries D78} (2008) 085011}
  [\href{https://arxiv.org/abs/0805.3993}{{\ttfamily 0805.3993}}].

\bibitem{Bern:2010ue}
Z.~Bern, J.~J.~M. Carrasco and H.~Johansson, \emph{{Perturbative Quantum
  Gravity as a Double Copy of Gauge Theory}},
  \href{https://doi.org/10.1103/PhysRevLett.105.061602}{\emph{Phys. Rev. Lett.}
  {\bfseries 105} (2010) 061602}
  [\href{https://arxiv.org/abs/1004.0476}{{\ttfamily 1004.0476}}].

\bibitem{Bern:2010yg}
Z.~Bern, T.~Dennen, Y.-t. Huang and M.~Kiermaier, \emph{{Gravity as the Square
  of Gauge Theory}},
  \href{https://doi.org/10.1103/PhysRevD.82.065003}{\emph{Phys. Rev.}
  {\bfseries D82} (2010) 065003}
  [\href{https://arxiv.org/abs/1004.0693}{{\ttfamily 1004.0693}}].

\bibitem{Bern:2013uka}
Z.~Bern, S.~Davies, T.~Dennen, A.~V. Smirnov and V.~A. Smirnov,
  \emph{{Ultraviolet Properties of N=4 Supergravity at Four Loops}},
  \href{https://doi.org/10.1103/PhysRevLett.111.231302}{\emph{Phys. Rev. Lett.}
  {\bfseries 111} (2013) 231302}
  [\href{https://arxiv.org/abs/1309.2498}{{\ttfamily 1309.2498}}].

\bibitem{Bern:2014sna}
Z.~Bern, S.~Davies and T.~Dennen, \emph{{Enhanced ultraviolet cancellations in
  $\mathcal N=5$ supergravity at four loops}},
  \href{https://doi.org/10.1103/PhysRevD.90.105011}{\emph{Phys. Rev.}
  {\bfseries D90} (2014) 105011}
  [\href{https://arxiv.org/abs/1409.3089}{{\ttfamily 1409.3089}}].

\bibitem{Bern:2017ucb}
Z.~Bern, J.~J.~M. Carrasco, W.-M. Chen, H.~Johansson, R.~Roiban and M.~Zeng,
  \emph{{Five-loop four-point integrand of $N=8$ supergravity as a generalized
  double copy}}, \href{https://doi.org/10.1103/PhysRevD.96.126012}{\emph{Phys.
  Rev.} {\bfseries D96} (2017) 126012}
  [\href{https://arxiv.org/abs/1708.06807}{{\ttfamily 1708.06807}}].

\bibitem{Bern:2018jmv}
Z.~Bern, J.~J. Carrasco, W.-M. Chen, A.~Edison, H.~Johansson, J.~Parra-Martinez
  et~al., \emph{{Ultraviolet Properties of $\mathcal N = 8$ Supergravity at
  Five Loops}}, \href{https://doi.org/10.1103/PhysRevD.98.086021}{\emph{Phys.
  Rev.} {\bfseries D98} (2018) 086021}
  [\href{https://arxiv.org/abs/1804.09311}{{\ttfamily 1804.09311}}].

\bibitem{Goldberger:2016iau}
W.~D. Goldberger and A.~K. Ridgway, \emph{{Radiation and the classical double
  copy for color charges}},
  \href{https://doi.org/10.1103/PhysRevD.95.125010}{\emph{Phys. Rev.}
  {\bfseries D95} (2017) 125010}
  [\href{https://arxiv.org/abs/1611.03493}{{\ttfamily 1611.03493}}].

\bibitem{Luna:2017dtq}
A.~Luna, I.~Nicholson, D.~O'Connell and C.~D. White, \emph{{Inelastic Black
  Hole Scattering from Charged Scalar Amplitudes}},
  \href{https://doi.org/10.1007/JHEP03(2018)044}{\emph{JHEP} {\bfseries 03}
  (2018) 044} [\href{https://arxiv.org/abs/1711.03901}{{\ttfamily
  1711.03901}}].

\bibitem{Goldberger:2017ogt}
W.~D. Goldberger, J.~Li and S.~G. Prabhu, \emph{{Spinning particles, axion
  radiation, and the classical double copy}},
  \href{https://doi.org/10.1103/PhysRevD.97.105018}{\emph{Phys. Rev.}
  {\bfseries D97} (2018) 105018}
  [\href{https://arxiv.org/abs/1712.09250}{{\ttfamily 1712.09250}}].

\bibitem{Shen:2018ebu}
C.-H. Shen, \emph{{Gravitational Radiation from Color-Kinematics Duality}},
  \href{https://doi.org/10.1007/JHEP11(2018)162}{\emph{JHEP} {\bfseries 11}
  (2018) 162} [\href{https://arxiv.org/abs/1806.07388}{{\ttfamily
  1806.07388}}].

\bibitem{Cheung:2018wkq}
C.~Cheung, I.~Z. Rothstein and M.~P. Solon, \emph{{From Scattering Amplitudes
  to Classical Potentials in the Post-Minkowskian Expansion}},
  \href{https://doi.org/10.1103/PhysRevLett.121.251101}{\emph{Phys. Rev. Lett.}
  {\bfseries 121} (2018) 251101}
  [\href{https://arxiv.org/abs/1808.02489}{{\ttfamily 1808.02489}}].

\bibitem{Bern:2019nnu}
Z.~Bern, C.~Cheung, R.~Roiban, C.-H. Shen, M.~P. Solon and M.~Zeng,
  \emph{{Scattering Amplitudes and the Conservative Hamiltonian for Binary
  Systems at Third Post-Minkowskian Order}},
  \href{https://arxiv.org/abs/1901.04424}{{\ttfamily 1901.04424}}.

\bibitem{Bahjat-Abbas:2017htu}
N.~Bahjat-Abbas, A.~Luna and C.~D. White, \emph{{The Kerr-Schild double copy in
  curved spacetime}},
  \href{https://doi.org/10.1007/JHEP12(2017)004}{\emph{JHEP} {\bfseries 12}
  (2017) 004} [\href{https://arxiv.org/abs/1710.01953}{{\ttfamily
  1710.01953}}].

\bibitem{Carrillo-Gonzalez:2017iyj}
M.~Carrillo-González, R.~Penco and M.~Trodden, \emph{{The classical double
  copy in maximally symmetric spacetimes}},
  \href{https://doi.org/10.1007/JHEP04(2018)028}{\emph{JHEP} {\bfseries 04}
  (2018) 028} [\href{https://arxiv.org/abs/1711.01296}{{\ttfamily
  1711.01296}}].

\end{thebibliography}

\providecommand{\href}[2]{#2}\begingroup\raggedright\endgroup

\end{document}